\documentstyle[aps,epsfig]{revtex} 
\begin{document} 
\draft 
\title{Multiboson effects in multiparticle production} 
\author{ 
R.~Lednicky$^{1,2}$, 
V.~Lyuboshitz$^{1,3}$, 
K.~Mikhailov$^{1,4}$, 
Yu.~Sinyukov$^{1,5}$,\\ 
A.~Stavinsky$^{1,4}$ 
and B.~Erazmus$^{1}$} 
\maketitle 
\maketitle 
\begin{center} 
{\small 
{\it 
 $^1$ 
SUBATECH, (UMR, Universite, 
Ecole des Mines, IN2P3/CNRS),\\ 
4, rue Alfred Kastler, F-44070 Nantes Cedex 03, France.\\ 
 $^2$ 
Institute of Physics, 
Na Slovance 2, 18221 Prague 8, Czech Republic.\\ 
 $^3$ 
JINR, 141980 Dubna, Moscow region, Russia.\\ 
 $^4$ 
ITEP, B.~Cheremushkinskaya 25, 117259 
Moscow, Russia.\\ 
 $^5$ 
ITP, Metrologicheskaya 14b, 252143  Kiev, Ukraine. 
}} 
\end{center} 
 
\begin{abstract} 
The influence of multiboson effects on pion multiplicities, 
single-pion spectra and two-pion correlation functions is discussed 
in terms of an analytically solvable model. 
The applicability of its basic factorization assumption is clarified. 
An approximate scaling of the basic observables
with the phase space density is demonstrated 
in the low density (gas) limit. This scaling and also its violation 
at high densities due to the condensate formation is described by 
approximate analytical formulae which allow, in principle, 
for the identification of the multiboson effects among others. 
For moderate densities indicated by the experimental data, 
a fast saturation of multiboson effects 
with the number of contributing cumulants is obtained, allowing for 
the account of these effects in realistic transport code simulations. 
At high densities, the spectra are mainly determined by the universal 
condensate term and the initially narrow Poisson 
multiplicity distribution approaches a wide Bose-Einstein one. 
%$\langle n(n-1)\rangle\rightarrow 2\langle n\rangle^2$. 
As a result, the intercepts of the inclusive and fixed-$n$ 
correlation functions (properly normalized to 1 at large 
relative momenta) 
approach 2 and 1, respectively and their widths logarithmically 
increase with the increasing phase space density. 
It is shown that the neglect of energy--momentum constraints in the 
model is justified except near a multipion threshold, where these 
constraints practically exclude the possibility of a very 
cold condensate production. 
It is argued that spectacular multiboson effects are likely to be observed 
only in the rare events containing sufficiently high density 
(speckle) fluctuations. 
\end{abstract} 
\section{Introduction}\label{sec1} 
In future heavy ion experiments at RHIC and LHC one expects to obtain 
thousands of pions per a unit rapidity interval. 
Since the pions are bosons 
there can be multiboson effects 
enhancing the production of pions with low relative momenta 
thus increasing the pion multiplicities, softening their spectra and 
modifying the correlation functions. 
Though the present data does not point to any spectacular 
multiboson effects, 
one can hope to observe new interesting 
phenomena like boson condensation or speckles in some rare 
events 
or in eventually overpopulated kinematic regions with the pion density 
in the 6-dimensional phase space, 
$f=(2\pi)^3 d^6n/d^3{\bf p}d^3{\bf x}$, 
of the order of unity 
(see, {\it e.g.}, 
\cite{pod89}-\cite{sl94}). 
%\cite{pod89,pod85,zaj87,zha93,pra93,pra94,ber94,sl94}). 
 
In the low-density limit ($f\ll 1$), the mean phase space density at a given 
momentum ${\bf p}$ can be estimated as the mean number of pions interfering 
with a pion of momentum ${\bf p}$ 
(rapidity $y$ and transverse momentum ${\bf p}_t$) 
and building the Bose-Einstein (BE) 
enhancement in the two-pion correlation function 
\cite{ber94,sl94}: 
$\langle f\rangle _{{\bf p}}\sim \pi^{3/2}{N}({\bf p})/V$, 
where ${N}({\bf p})=d^3{n}/d^3{\bf p}$ and 
$V=r_xr_yr_z$ is the interference volume defined in terms of the 
outward ($r_x$), sideward ($r_y$) and longitudinal ($r_z$) 
interferometry radii. 
Typically $\langle f\rangle _{{\bf p}}\sim 0.1$ 
for mid-rapidities and $p_t\sim \langle p_t\rangle$ \cite{ber94}. 
The data are also consistent with the phase space density of pions 
near the local thermal equilibrium \cite{bar97}. 
 
At AGS and SPS energies the interference volume $V$ seems to 
scale with $dn/dy$ 
(see, {\it e.g.}, \cite{jac94,gaz94}) 
pointing to the freeze-out of the pions 
at a constant phase space density.\footnote 
{Similar effect was observed also for protons produced in hadron- and 
electron-nucleus interactions \cite{sta94}. 
} 
If this trend will survive 
then there will be no spectacular multiboson effects 
in the ordinary events at RHIC or even at LHC. 
In such a situation the standard 
two-particle interferometry technique could be used 
to measure the space-time intervals between the production points 
also in the future collider experiments. 
The corresponding interferometry radii for lead-lead collisions at 
LHC would be however rather large - about 20 fm.

The multiboson effects can show up however in certain classes 
of events. 
An example is a rapidly expanding system with the entropy much smaller 
than in the case of total equilibrium. 
Then a strong transverse 
flow can lead to rather dense gas of soft pions in the central part of the 
hydrodynamic tube at the final expansion stage 
(see, {\it e.g.}, \cite{akk95}). 
Another reason can be the formation of quark-gluon 
plasma or mixed phase. 
Due to large gradients of temperature 
or velocity the hydrodynamic layer near the boundary with vacuum can 
decay at a large phase space density and lead to pion speckles even at 
moderate transverse momenta \cite{sla91}. 
 
The dramatic difference in behavior of Boltzman-like gases and dense 
multiboson systems can lead to serious problems for transport models like 
RQMD, VENUS, {\it etc.}, ignoring actually the statistical properties 
of the particles both in intermediate and final states. In these models the 
most intensive particle production happens at relatively early evolution 
(expansion) stage when rather large pion phase space densities can be 
achieved at RHIC or LHC energies. 
 
Generally, the account of the multiboson effects is extremely difficult task. 
Even on the neglect of particle interaction in the final state the 
requirement of the BE symmetrization leads to severe numerical problems 
which increase factorially with the number of produced bosons 
\cite{zaj87,zha93}. 
In such a situation, it is important that there exists a simple analytically 
solvable model \cite{pra93} allowing for a study of the characteristic 
features of the multiboson systems under various conditions including 
those near the Bose condensation. 
In this paper we use this model to demonstrate the influence of the 
multiboson effects on pion multiplicities, 
spectra and two-pion correlation functions. 
Besides the original papers \cite{pra93,pra94}, similar studies can be 
found also in \cite{alinote95,cha95,zim97}. 
Particularly, some of the new aspects of the 
multiboson effects, like the scaling behavior with the phase 
space density or the behavior of the (semi-)inclusive correlation functions 
near the condensation limit were studied in our unpublished paper 
\cite{alinote95}. The present work represents an elaborated version 
of the latter. 
 
In Section \ref{sec2} we introduce the space-time description of particle 
production in terms of Wigner-like densities and discuss their 
physical meaning and the conditions of their factorization in the 
model of classical one-particle sources. 
The multiboson formalism in the factorizable case is set forth 
in Section \ref{sec3}. 
Using this formalism and the simple Gaussian ansatz for the emission 
function, we present in Section \ref{sec4} the analytical solutions 
(partly in terms of the recurrence relations) for the 
multiplicity distribution, single-boson spectra and two-boson 
correlation functions. 
In Section \ref{sec5} we compare the results of numerical 
calculations with the 
analytical approximations accounting for the approximate scaling 
behavior in the low density (gas) limit as well as for the condensate 
formation at high densities. 
The results are discussed and summarized in Section \ref{sec6}.

\section{Space-time picture of particle production}\label{sec2} 
\subsection{Wigner-like density}\label{sec2a} 
Let us first consider a process in which, 
besides others, just $n$ non-identical particles of given types 
are produced with the $4$-momenta 
$p_i=\{E_i,{\bf p}_i\}$ and Lorentz factors $\gamma_i=E_i/m_i$ 
(to simplify the notation, we assume that particles are spin--less). 
The inclusive differential cross section of this process is described by the 
invariant production amplitude 
$T_n(p_1,\dots,p_n;\alpha )$: 
\begin{equation} 
\label{1A}\gamma_1\cdots\gamma_n\frac{d^{3n}\sigma _n} 
{d^3{\bf p}_1\cdots d^3{\bf p}_n}= 
\sum\limits_{\alpha }\mid T_n(p_1,\dots,p_n;\alpha )\mid ^2 
\equiv \gamma_1\cdots\gamma_n\sigma _n P_n({\bf p}_1,\dots,{\bf p}_n), 
\end{equation} 
where the sum over the quantum numbers $\alpha$, 
describing the rest of the produced system, contains also an 
integration over the momenta of the other produced particles with the 
energy-momentum conservation taken into account. The 
non-invariant production probability $P_n({\bf p}_1,\dots,{\bf p}_n)$ 
is normalized to unity. 

If the particles are identical spin--less bosons, then the production 
amplitude has to satisfy the requirement of Bose symmetry. 
Formally, this can be achieved by the substitution of the 
non-symmetrized amplitude $T_n(p_1,\dots,p_n;\alpha )$ 
(corresponding to the "switched off" effect of quantum statistics) 
by a properly symmetrized one: 
\begin{equation} 
\label{1B} T_n(p_1,\dots,p_n;\alpha )\rightarrow (n!)^{-1/2} 
\sum\limits_{\sigma}T_n(p_{\sigma_1},\dots,p_{\sigma_n};\alpha ), 
\end{equation} 
where the sum is over all $n!$ permutations $\sigma $ of 
the sequence $\{1,2,\ldots n\}$. 
 
In the following we will neglect particle interaction in 
the final state.\footnote 
{This is more or less valid assumption for neutral pions but not 
for the charged ones. For the treatment of multiboson effects in the case 
of interacting pions see, {\it e.g.}, \cite{ame-led}.} 
Then the 
non-symmetrized amplitude 
$T_n(p_1,\dots,p_n;\alpha )\equiv T_n({\mathcal P};\alpha)$ is related to the 
amplitude in the space-time representation 
${\mathcal T}_n(x_1,\dots,x_n;\alpha )
\equiv {\mathcal T}_n({\mathcal X};\alpha)$ 
(describing production of the 
particles in the space-time points with the 4-coordinates 
${\mathcal X}=\{x_1,\dots,x_n\}$) 
by the usual Fourier transform: 
\begin{equation} 
\label{2A} T_n({\mathcal P};\alpha )=\int d{\mathcal X} {\rm e}^{ 
i{\mathcal P}{\mathcal X}}{\mathcal T}_n({\mathcal X};\alpha ). 
\end{equation} 
Inserting (\ref{2A}) into (\ref{1A}) and introducing the space-time 
density matrix \cite{ll90}: 
$
\rho _n({\mathcal X};{\mathcal X}^{\prime })= 
\sum\limits_{\alpha }{\mathcal T}_n({\mathcal X};\alpha ) 
{\mathcal T}_n^{*}({\mathcal X}^{\prime };\alpha ) 
$
and its partial Fourier transform (emission function) - 
an analogy of the Wigner density 
%\cite{shu73,PRA84}: 
\cite{PRA84}: 
\begin{equation} 
\label{17A} 
\begin{array}{c} 
\widetilde{D}_n(p_1,\bar x_1;\dots;p_n,\bar x_n)\equiv 
\gamma_1\cdots\gamma_n\sigma_n{D}_n(p_1,\bar x_1;\dots;p_n,\bar x_n) 
%\\ \\ 
=\int d\delta {\rm e}^{i{\mathcal P}\delta}  
\rho_n(\bar{\mathcal X}+\frac 12\delta; 
\bar{\mathcal X}-\frac12\delta), 
\end{array} 
\end{equation} 
where 
$\bar{\mathcal X} =\frac 12({\mathcal X}+{\mathcal X}^{\prime })$, 
$\delta ={\mathcal X}-{\mathcal X}^{\prime }$, 
we can rewrite the production cross section (\ref{1A}) in the absence 
of BE effects as: 
\begin{equation} 
\label{1AA}\frac{d^{3n}\sigma _n} 
{d^3{\bf p}_1\cdots d^3{\bf p}_n}\equiv 
\sigma_n P_n({\bf p}_1,\dots,{\bf p}_n) 
=\sigma_n\int d^4\bar x_1\cdots d^4\bar x_n 
{D}_n(p_1,\bar x_1;\dots;p_n,\bar x_n). 
\end{equation} 
Since ${D}_n$ 
is a real (though not positively defined) 
function normalized to unity, in accordance with 
Eq.~(\ref{1AA}) it can be considered as 
an approximation to the emission probability of the particles with given 
4-momenta $p_i$ in the average space-time points 
$\bar x_i=\frac 12(x_i+x_i^{\prime })$. 
 
The insertion of the symmetrized amplitude (\ref{1B}) into the 
cross section formula (\ref{1A}) leads to the substitution of the 
probability $P_n$ by a BE modified one $P_n^c$. 
For example, in case of a two-boson production process, 
instead of Eq. (\ref{1AA}) we have: 
\begin{equation} 
\label{Pc2} 
P_2^c({\bf p}_1,{\bf p}_2)=\int d^4x_1 d^4x_2[ 
D_2(p_1,x_1;p_2,x_2)+D_2(p,x_1;p,x_2)\cos(qx)], 
\end{equation} 
where $p=\frac12(p_1+p_2)$, $q=p_1-p_2$ and $x=x_1-x_2$. 
Clearly, for $n>1$ the probability $P_n^c$ is no more normalized to unity. 
The integral over this probability 
yields the BE weight $\omega_n$ of an $n$-boson event 
produced in the absence of the effect of quantum statistics 
($\omega_0=\omega_1=1$, $\omega_n>1$ for $n>1$): 
\begin{equation} 
\label{normc} \omega_n =
\int d^3{\bf p}_1\cdots d^3{\bf p}_n 
P_n^c({\bf p}_1,\dots,{\bf p}_n)/
\int d^3{\bf p}_1\cdots d^3{\bf p}_n 
P_n({\bf p}_1,\dots,{\bf p}_n). 
\end{equation} 
We will also use the differential 
BE weights $\omega_n^{(k)}({\bf p}_1,\dots,{\bf p}_k)$ defined as 
in Eq. (\ref{normc}) except for a skipped integration over the momenta
of particles $1,2,\dots,k$.

The multiboson problem greatly simplifies (see Section \ref{sec3}) 
in the factorizable case when the 
$n$-particle emission function and, as a consequence, the 
non-symmetrized production probability can be written as 
products of the single-particle ones: 
\begin{equation} 
D_n(p_1,x_1;\dots;p_n,x_n)=D(p_1,x_1)\cdots D(p_n,x_n),~~~ 
P_n({\bf p}_1,\dots,{\bf p}_n)=P({\bf p}_1)\cdots P({\bf p}_n). 
\label{8.74a} 
\end{equation} 
Consequently, the BE weights are expressed through a universal 
function $F_{ij}$ (see, {\it e.g.}, \cite{ame-led}): 
\begin{equation} 
\label{onn}\omega_n^{(n)}({\bf p}_1,{\bf p}_2,\ldots {\bf p}_n)= 
\sum\limits_\sigma 
\prod\limits_{i=1}^nF_{i\sigma _i}, 
\end{equation} 
where 
\begin{equation} 
\label{Fij} 
F_{ij}=\int d^4xD(p_{ij},x)\cdot \exp(iq_{ij}x)\left/\right. 
[P({\bf p}_i)P({\bf p}_j)]^{1/2}, 
\end{equation} 
$p_{ij}=\frac12(p_i+p_j)$ and $q_{ij}=p_i-p_j$. 
The sum in Eq. (\ref{onn}) is over $n!$ possible permutations $\sigma $ of 
the sequence $\{1,2,\ldots n\}$. 
For example, in the two-boson case, we have 
\begin{equation} 
\label{o2}\omega_2^{(2)}({\bf p}_1,{\bf p}_2)= 
F_{11}F_{22}+F_{12}F_{21}\equiv 1+K_2^{(2)}({\bf p}_1,{\bf p}_2), 
\end{equation} 
where $K_2^{(2)}=F_{12}F_{21}$ 
is so called differential cumulant (see Section \ref{sec3}). 
 
\subsection{One-particle sources}\label{sec2b} 
To clarify the physical meaning of the emission function and the 
factorization assumption (\ref{8.74a}), let us follow 
Kopylov and Podgoretsky (see, {\it e.g.}, \cite{pod89}) 
and assume that particles $1, 2,\dots$ are emitted by one-particle sources 
$A, B,\dots$ which are considered as classical so 
they can be treated 
by parameters and not by amplitudes. 
Thus the $4$-coordinates of the source 
centers $x_A,x_B,\dots$ and other source characteristics in the model can be 
considered as a part of the quantum numbers $\alpha \equiv 
\{x_A,x_B,\dots\alpha ^{\prime }\}$. It was pointed out by Kopylov and 
Podgoretsky that the BE effect is mainly determined by the phase factor 
exp$(ip_1x_A+ip_2x_B+\cdots)$ contained in the amplitude 
$T_n(p_1,\dots,p_n;\alpha )$. 
 
Let us first consider the production of only one boson. 
Assuming the translation invariance of the decay amplitudes 
$\widetilde{u}$, we can write 
the single-boson amplitude in the 4-coordinate representation as: 
\begin{equation} 
\label{3A}{\mathcal T}(x_1;\alpha )\equiv 
\widetilde{t}(x_1-x_A;\alpha )= 
\widetilde{u}(x_1-x_A;\alpha ^{\prime })v(\alpha), 
\end{equation} 
where $\alpha=\{x_A,\alpha ^{\prime }\}$ and 
$\alpha ^{\prime }=\{A,\dots \}$. 
Inserting (\ref{3A}) into (\ref{2A}) and introducing the Fourier transform 
\begin{equation} 
\label{4A}t(p;\alpha )=\int d^4\xi {\rm e}^{ip\xi} 
\tilde t(\xi ;\alpha )\equiv 
u(p;\alpha ^{\prime })v(\alpha), 
\end{equation} 
we obtain the Kopylov-Podgoretsky ansatz: 
\begin{equation} 
\label{5}T(p;\alpha )={\rm e}^{ip x_A}t(p;\alpha)\equiv 
{\rm e}^{ip x_A}u(p;\alpha ^{\prime })v(\alpha). 
\end{equation} 
For the production probability we have 
\begin{equation} 
\label{Pc1a} 
P^c({\bf p})=P({\bf p})\equiv \int d^4x_A W(p,x_A) 
=\int d^4x_A\sum\limits_{\alpha ^{\prime }}\mid 
t(p;\{x_A,\alpha ^{\prime }\})\mid ^2, 
\end{equation} 
where we have introduced the (true) emission probability $W({\bf p},x_A)$. 
Similarly, if two bosons are produced, the production probability 
takes on the form: 
\begin{equation} 
\label{8A} 
\begin{array}{c} 
P_2^{ 
{c}}({\bf p}_1,{\bf p}_2)= \int d^4x_Ad^4x_B\sum\limits_{\alpha 
^{\prime }}\left\{ \mid t(p_1,p_2;\alpha )\mid ^2+ 
Re[t(p_1,p_2;\alpha 
)t^{*}(p_2,p_1;\alpha ){\rm e}^{ iqx}]\right\}. 
\end{array} 
\end{equation} 
Note that here $x=x_A-x_B$ and, as usual, $q=p_1-p_2$. 
 
The emission function $D$ can be expressed through the 
Kopylov-Podgoretsky amplitudes $t(p;\alpha )$ continued off 
mass-shell. Using the 
inverted Fourier transform in Eq. (\ref{4A}), we get from Eq. (\ref{17A}) 
\cite{ame-led}: 
\begin{equation} 
\label{19A} 
\begin{array}{c} 
D(p,\bar x_1)=\frac 1{(2\pi )^4}\sum\limits_{\alpha 
^{\prime }}\int d^4x_A d^4\kappa {\rm e}^{i\kappa 
(x_A-\bar x_1)} 
\cdot t(p+\frac 12\kappa ;\{x_A,\alpha ^{\prime }\}) 
t^{*}(p-\frac 12\kappa ;\{x_A,\alpha ^{\prime }\}) 
\nonumber \\ 
=\frac 1{(2\pi )^4}\sum\limits_{\alpha 
^{\prime }}\int d^4x_A d^4\kappa {\rm e}^{i\kappa 
(x_A-\bar x_1)} 
\cdot u(p+\frac 12\kappa ;\alpha ^{\prime }) 
u^{*}(p-\frac 12\kappa ;\alpha ^{\prime }) 
|v(\alpha)|^2. 
\end{array} 
\end{equation}

It is clear from Eq. (\ref{4A}) that the momentum dependence of the amplitude 
$t(p;\alpha )$ is determined by the space-time extent of the 
one-particle source. For example, assuming that the source emits a particle 
independently of the quantum numbers $\alpha ^{\prime }$ 
except for the source type $A$: 
$\widetilde{u}(\xi ;\alpha ^{\prime })=\widetilde{u}(\xi;A)$ 
and that the distribution of the emission points in the source rest frame is 
given by a simple Gaussian with
the width parameters $r_A$ and $\tau _A$ characterizing the proper 
space-time sizes of the source $A$, 
we obtain in case of a source at rest:\footnote 
{Note that Eq. (\ref{7A}) is valid also off mass-shell when $% 
p_{0}\neq E$. 
For a source moving with a non-relativistic velocity 
$\mbox{\boldmath$\beta$}_A$, the substitution 
${\bf p}\rightarrow {\bf p}-{\bf p}_A$ 
has to be done in Eq. (\ref{7A}), where 
${\bf p}_A=m\mbox{\boldmath$\beta$}_A$ is a mean 
3-momentum of the particle emitted by a source $A$. 
} 
\begin{equation} 
\label{7A}t(p;\alpha )\sim \exp (-\frac 12r_A^2{\bf p}^2-\frac 
12\tau _A^2p_{0}^2)v(\alpha ). 
\end{equation} 
The probability $\left| v(\alpha )\right| ^2$ describes the production 
of particle sources and depends on the $4$-coordinates 
$x_A$ of the source centers. 
In the following we will take it also in a simple Gaussian form
with the width parameters $\widetilde{r}_0$ and $\widetilde{\tau }_0$ 
characterizing the space-time extent of the source production region. 
 
Comparing Eqs.~(\ref{Pc1a})  and (\ref{19A}), it can be seen that the emission 
function $D$ is more spread in space and time than the emission 
probability $W$. In particular, the Gaussian parameterizations 
of $\widetilde{u}(\xi;A)$ and $\left| v(\alpha )\right| ^2$ yield: 
\begin{equation} 
\label{20A}
W(p,x)\sim \exp (-r_A^2{\bf p}^2-\tau _A^2p_0^2)\exp (-\frac{{\bf x% 
}^2}{2\widetilde{r}_0^2}-\frac{x_0^2}{2\widetilde{\tau }_0^2}) 
\end{equation} 
and 
\begin{equation} 
\label{21A} 
D(p,x)\sim \exp (-r_A^2{\bf p}^2-\tau _A^2p_0^2)\exp (-\frac{{\bf x% 
}^2}{2\widetilde{r}_0^2+r_A^2}-\frac{x_0^2}{2\widetilde{\tau }_0^2+\tau _A^2}). 
\end{equation} 
 
Clearly, the factorized form (\ref{8.74a}) for the emission 
function is recovered in case of independent sources 
({\it i.e.} sources having no quantum numbers in common) 
assuming a unique mechanism of their production. 
Generally, the latter condition may not be fulfilled, 
{\it e.g.} in case of heavy ion collisions 
without selection of the impact parameter. 
Then, even for independent sources, Eq. (\ref{8.74a}) will be 
substituted by a weighted sum of factorized terms corresponding to 
different single-particle emission functions. 
 
The single-particle density generally contains contributions from 
the sources of different type ({\it e.g.} different resonances). 
It is interesting to note that in case of only one source type 
({\it i.e.} universal source parameters) and on condition 
of sufficiently slow relative motion of the sources contributing to 
low-$|q|$ pairs ({\it e.g.}, due to limited source decay momentum), 
the BE correlation effect in the two-boson case is solely 
determined by the characteristic space-time distance between the 
source centers: 
\begin{equation} 
\label{p2g} 
P_2^{{c}}({\bf p}_1,{\bf p}_2)=P({\bf p}_1)P({\bf p}_2)[1+ 
\exp (-\widetilde{r}_0^2{\bf q}^2-\widetilde{\tau }_0^2 q_0^2)]. 
\end{equation} 
This result follows from Eq. (\ref{Pc2}) or, in terms of the 
Kopylov-Podgoretsky emission amplitudes, immediately from 
Eq. (\ref{8A}). If the sources of interfering bosons 
move with non-relativistic 
velocities $\mbox{\boldmath$\beta$}_A\equiv {\bf p}_A/m$ and the 
distribution of their characteristics is given by a product of
$\exp(-{\bf p}_A^2/2\Delta_0^2)$ and
the same Gaussian as before 
({\it i.e.} the residual slow relative motion decouples from other 
source characteristics), we still arrive at 
Eqs. (\ref{20A}), (\ref{21A}), up to a substitution 
$r_A^2{\bf p}^2\rightarrow r_A^2{\bf p}^2/[2(r_A\Delta_0)^2+1]$ 
corresponding to a widening of the momentum distribution 
due to the source relative motion. 
Note that the factorized form of the multiparticle 
density is not destroyed by this motion. The latter however 
influences the two-boson correlation function which now 
becomes sensitive to the source size $r_A$ even in case 
of one source type, Eq. (\ref{p2g}) being modified by the 
substitution 
$\widetilde{r}_0^2\rightarrow  \widetilde{r}_0^2  + 
r_A^2/[2+(r_A\Delta_0)^{-2}]$. 
 
The space-time extent of one-particle sources can be usually considered 
much smaller than the characteristic space-time distance between their 
centers ($r_A\ll \widetilde{r}_0$, $\tau_A\ll \widetilde{\tau}_0$). 
The 4-momentum dependence of one-particle amplitudes is then 
negligible when varying the particle 4-momenta by the amount 
$\sim \widetilde{r}_0^{-1}, \widetilde{\tau}_0^{-1}$ characteristic for the 
interference effect. On such a {\it smoothness} condition, there is 
practically no difference between the emission function $D$ and the 
emission probability $W$ and both Eqs.~(\ref{Pc2}) and (\ref{8A}) 
yield the well known result of Kopylov-Podgoretsky 
for the production probability of two identical bosons: 
\begin{equation} 
\label{p2gG} 
P_2^{{c}}({\bf p}_1,{\bf p}_2)\doteq P({\bf p}_1,{\bf p}_2)[1+ 
\langle\cos(q(x_1-x_2))\rangle]. 
\end{equation}

\section{Multiboson formalism in factorizable case}\label{sec3} 
 
The multiboson effects can be practically treated provided that we can 
neglect particle interaction in the final state and assume independent 
emission of non-interfering particles 
(a valid assumption for heavy ion collisions), 
supplemented by the requirement of a universal single-particle 
emission function $D(p,x)$ for the detected class of events. 
We can then use Eq.~(\ref{8.74a}) expressing 
the $n$-particle emission function as 
a product of the single-particle ones. 
Then, similar to refs. \cite{pra93,pra94} it is convenient to define 
the functions 
\begin{eqnarray} 
G_1({\bf p}_1,{\bf p}_2) &=& 
\int d^4xD({\textstyle \frac 12 }(p_1+p_2),x)\cdot \exp(i(p_1-p_2)x), 
\nonumber \\ 
G_n({\bf p}_1,{\bf p}_2)&=& 
\int d^3{\bf k}_2\dots d^3{\bf k}_n G_1({\bf p}_1,{\bf k}_2)\dots 
G_1({\bf k}_n,{\bf p}_2) 
\nonumber \\ 
&\equiv& 
\int d^3{\bf k}_2 G_{n-1}({\bf p}_1,{\bf k}_2) 
G_1({\bf k}_2,{\bf p}_2), 
\nonumber \\ 
g_n&=&\int d^3{\bf p} G_{n}({\bf p},{\bf p}). 
\label{8.9} 
\end{eqnarray} 
The function $G_1$ at equal momenta is just 
the initial (not affected by the multiboson effects) 
single-boson spectrum normalized to unity: 
\begin{equation} 
P({\bf p})=G_1({\bf p},{\bf p}),~~ 
\int d^3{\bf p} P({\bf p}) \equiv g_1=1. 
\label{8.130} 
\end{equation} 
The related quantities are so called cumulants 
\begin{eqnarray} 
K_n^{(2)}({\bf p}_1,{\bf p}_2) &=& (n-2)! 
\sum_{i=1}^{n-1}G_i({\bf p}_1,{\bf p}_2) 
G_{n-i}({\bf p}_2,{\bf p}_1)\left/\right. 
[P({\bf p}_1)P({\bf p}_2)], 
\nonumber \\ 
K_n^{(1)}({\bf p})&=& (n-1)!G_n({\bf p},{\bf p})\left/\right.P({\bf p}), 
\nonumber \\ 
K_n&=& (n-1)!g_n. 
\label{8.131} 
\end{eqnarray} 
It can be shown that the BE weight of an event with 
n identical spin-zero bosons is determined through the cumulants $K_j$ 
by the recurrence relation \cite{ame-led}: 
\begin{equation} 
\omega_n= C_0^{n-1}K_1\omega_{n-1}+C_1^{n-1}K_2\omega_{n-2}+ 
\dots +C_{n-1}^{n-1}K_n\omega_0 
%\rightarrow c(\beta)n!/\beta^n 
\label{8.10} 
\end{equation} 
with $\omega_0=\omega_1=1$; 
$C_i^{n-1}=(n-1)!/[i!(n-1-i)!]$ are the usual combinatorial numbers. 
For example, $\omega_2=1+K_2$ and $\omega_3=1+3K_2+K_3$. 
One can check that $\omega_n=n!$ provided that all the elementary 
one-particle sources are situated at one and the same space-time point so that 
all the single-boson states are identical and 
$K_{j+1}=j!$.\footnote 
{This situation is similar (flat correlation function) 
though different from the case of the emission of so called 
coherent bosons for which there is no enhancement factor. 
In fact, when the one-particle sources become closer and closer, 
so that  their distances are less than the wave length of the emitted bosons, 
they can no more be considered as independent ones and a 
multiparticle source of non-interfering bosons has to be introduced 
\cite{LLP83}. To quantify the transition to the non-interfering bosons 
a concept of the coherence length can be used \cite{ST90}. 
} 
In the other extreme case of a large phase space volume and 
$n^2K_2\ll 1$, we can neglect the contribution of the 
higher order cumulants except for 
the first power of $K_2$ and write: 
\begin{equation}
\omega_n\doteq 1 + C_2^{n}K_2. 
\label{oappr} 
\end{equation}

Given the initial multiplicity distribution $\widetilde{w}(n)$, 
the BE affected one is easily calculated using the BE weights $\omega_n$: 
\begin{equation} 
w(n)= \omega_n \widetilde{w}(n)\left/\right. 
\sum_{j=0}^{\infty}\omega_j\widetilde{w}(j). 
\label{mulBE} 
\end{equation} 
Particularly, assuming the initial Poissonian distribution 
with the mean multiplicity $\eta$: 
$\widetilde{w}(n)={\rm e}^{-\eta}\eta^n/n!$, 
we get: 
\begin{equation} 
w(n)= \omega_n\frac{\eta^n}{n!}\left/\right.\sum_{j=0}^{\infty} 
\omega_j\frac{\eta^j}{j!} . 
\label{mult1} 
\end{equation} 
Similarly, the BE affected single- and two-boson spectra, respectively 
normalized to $n$ and $n(n-1)$, can be written as 
\begin{equation} 
{N}_n^{(1)}({\bf p})\equiv {N}_n({\bf p}) = n \omega_n^{(1)}({\bf p}) 
\cdot P({\bf p})\left/\right.\omega_n 
\label{8.12} 
\end{equation} 
and 
\begin{equation} 
{N}_n^{(2)}({\bf p}_1,{\bf p}_2)= n(n-1) 
\omega_n^{(2)}({\bf p}_1,{\bf p}_2)P({\bf p}_1)P({\bf p}_2) 
\left/\right.\omega_n, 
\label{8.13} 
\end{equation} 
where the differential BE weights $\omega_n^{(1)}({\bf p})$ and 
$\omega_n^{(2)}({\bf p}_1,{\bf p}_2)$ 
are expressed through the differential cumulants 
$K_n^{(1)}({\bf p})$ and $K_n^{(2)}({\bf p}_1,{\bf p}_2)$ \cite{alinote95}: 
\begin{eqnarray} 
\omega_n^{(1)}({\bf p})\equiv  \int d^3{\bf p}' 
\omega_n^{(2)}({\bf p},{\bf p}') 
P({\bf p}') 
= \sum_{j=0}^{n-1}C_j^{n-1}K_{j+1}^{(1)}({\bf p}) 
\omega_{n-1-j}, 
\nonumber \\ 
\omega_n^{(2)}({\bf p}_1,{\bf p}_2) = \sum_{j=0}^{n-2}C_j^{n-2}\omega_{n-2-j} 
\left[\sum_{l=0}^jC_l^j K_{l+1}^{(1)}({\bf p}_1)K_{j-l+1}^{(1)}({\bf p}_2)+ 
K_{j+2}^{(2)}({\bf p}_1,{\bf p}_2)\right]. 
\label{cf1} 
\end{eqnarray} 
 
The differential weight $\omega_n^{(2)}({\bf p}_1,{\bf p}_2)$ 
can be considered as a two-particle correlation function 
measuring the BE effect on the initial uncorrelated 
two-particle spectrum 
$\widetilde{{N}}_n^{(2)}({\bf p}_1,{\bf p}_2)= 
n(n-1)P({\bf p}_1)P({\bf p}_2)$, 
with the normalization 
\begin{equation} 
\int d^3{\bf p}_1d^3{\bf p}_2 \omega_n^{(2)}({\bf p}_1,{\bf p}_2) 
P({\bf p}_1)P({\bf p}_2)=\omega_n. 
\end{equation} 
Usually the correlation function is normalized to unity at a large 
$|{\bf q}|$. 
Such a normalization is approximately satisfied for the 
correlation function defined as: 
\begin{equation} 
R_n({\bf p}_1,{\bf p}_2)= {N}_n^{(2)}({\bf p}_1,{\bf p}_2)\left/\right. 
\widetilde{{N}}_n^{(2)}({\bf p}_1,{\bf p}_2)\equiv 
\omega_n^{(2)}({\bf p}_1,{\bf p}_2)\left/\right.\omega_n. 
\label{cfa2} 
\end{equation} 
In practice, the two-particle correlation function is defined through the 
observable spectra as: 
\begin{equation} 
R_n({\bf p}_1,{\bf p}_2)= c_n{N}_n^{(2)}({\bf p}_1,{\bf p}_2)\left/\right. 
[{N}_n^{(1)}({\bf p}_1){N}_n^{(1)}({\bf p}_2)]. 
\label{cf2} 
\end{equation} 
Similarly, the (semi-)inclusive correlation function is defined as 
\begin{equation} 
R({\bf p}_1,{\bf p}_2)= c~{N}^{(2)}({\bf p}_1,{\bf p}_2)\left/\right. 
[{N}^{(1)}({\bf p}_1){N}^{(1)}({\bf p}_2)], 
\label{cf2i} 
\end{equation} 
where 
\begin{equation} 
{N}^{(1)}({\bf p})= 
\sum_n w(n){N}_n^{(1)}({\bf p}),~~~~ 
{N}^{(2)}({\bf p}_1,{\bf p}_2)= 
\sum_n w(n){N}_n^{(2)}({\bf p}_1,{\bf p}_2) 
\end{equation} 
are the corresponding (semi-)inclusive single- and two-particle spectra, 
$w(n)$ is the normalized multiplicity distribution accounting for the 
BE effect according to Eq. (\ref{mulBE}). 
Later on, using an analytical Gaussian model for the emission function, 
we show that the normalization constant $c_n$ can be 
expressed through the BE weights as: 
\begin{equation} 
c_n=  n\omega_{n-1}^2\left/\right.[(n-1)\omega_n\omega_{n-2}] 
\label{c_n} 
\end{equation} 
and that $c=1$ for the inclusive correlation function provided a 
Poissonian multiplicity distribution of the initially uncorrelated bosons. 
 
As one can see from formulae (\ref{8.12})-(\ref{cf1}), the multiboson 
correlations lead to distortions of the initial 
single- and two-particle distributions. Such distortions are small in the 
case of interference of only two or three identical particles. However, 
they can become essential for the events with a large number of 
identical bosons due to factorially increasing number of the 
correction terms \cite{zaj87} (see also \cite{pod89} and \cite{pra93}). 
For the processes characterized by 
a high ($> 0.1$) phase space density of the identical bosons at the 
freeze-out time the  multiboson effects can no more be considered 
as a correction \cite{zaj87}.

To account for the multiboson symmetrization effect in the event 
simulators, 
a phase space weighting procedure was used with 
weights in the form of a normalized square of the sum of $n!$ plane waves 
\cite{zaj87,zha93}. 
This procedure however 
appears not practical for a large $n$ due to the 
factorially large number of the terms to be computed to calculate the weight 
and, due to large weight fluctuations. These fluctuations can be 
substantially reduced by weighting only in the momentum space. The 
corresponding BE weights are given in Eq. (\ref{onn}). 
They are expressed through the universal function (\ref{Fij}) 
which is simply related with the function $G_1$: 
\begin{equation} 
F_{ij}=G_1({\bf p}_i,{\bf p}_j)\left/\right.[P({\bf p}_i)P({\bf p}_j)]^{1/2}. 
\end{equation} 
On the condition of sufficient smoothness of the single-particle spectra, 
we can put 
\begin{equation} 
F_{ij}\doteq 
\int d^4xD(p_{ij},x)\cdot \exp(iq_{ij}x)\left/\right. 
\int d^4xD(p_{ij},x), 
\end{equation} 
where $p_{ij}=\frac12(p_i+p_j)$ and $q_{ij}=p_i-p_j$. 
This function can then be calculated as suggested in \cite{ame-led}: 
\begin{equation} 
\label{24d} 
F_{ij}=\left\langle \exp (iq_{ij}x_{k})\right\rangle _{{\bf p}_{ij}}, 
\end{equation} 
where the averaging is done over all simulated phase space points 
$\{p_k,x_k\}$ such that ${\bf p}_k$ is close to a given 3-momentum 
${\bf p}_{ij}$. 
However, there is still the problem 
with factorially large number of the terms 
required to calculate the weight 
according to Eq. (\ref{onn}). 
 
Fortunately, when calculating 
only single- or two-particle distributions according to Eqs. (\ref{8.12}) or 
(\ref{8.13}), 
this number is strongly reduced (eaten by the combinatorial numbers 
$C_j^m$ in Eqs. (\ref{cf1})). 
We should however perform integration 
over momenta of one or more particles to determine the integrated cumulants 
$K_n^{(2)}({\bf p}_1,{\bf p}_2)$, 
$K_n^{(1)}({\bf p})$ and $K_n$. 
 
The numerical averaging of the 
cumulants of all orders is a difficult task. 
In case of large multiplicities of identical bosons ($n > 20$) 
this is practically possible in the models with a symmetric emission 
function (allowing to use a special Monte Carlo technique) \cite{zaj87} 
or with a simple analytical parameterization of this function 
\cite{pra93,pra94}. For example, in ref. \cite{pra94} 
the corrections to multiplicity distributions, 
single-particle spectra and two-particle correlation functions were 
calculated using the relativistic Bjorken model \cite{Bjor83} for the 
emission function. To compute cumulants up to tenth order, the integration 
was performed analytically over the space-time coordinates and numerically 
over the momenta. 
 
Generally, for realistic models used to predict particle production in 
ultra-relativistic heavy-ion collisions, the numerical 
limitations allow to determine only a few lowest order cumulants 
(up to about the fourth order) \cite{ame-led}. 
Fortunately, since the 
interferometry measurements point out to a moderate 
pion freeze-out phase space 
density of $\sim 0.1$, 
the lowest order cumulant approximation appears to be reasonable 
for typical events in present and likely also in future 
heavy-ion experiments (see Section \ref{sec5}). 
At the same time, even in the absence of strong multiboson effects, 
their account can still be 
important for realistic simulations of heavy ion 
collisions \cite{ame-led}.

\section{Analytical model}\label{sec4} 
 
To study the multiboson effects in a dense pion gas, we 
use a simple model 
assuming independent particle emission (see Eq. (\ref{8.74a})) 
with the Gaussian ansatz for the single-boson emission 
function $D(p,x)$ \cite{pra93}: 
\begin{equation} 
D(p,x)=\frac{1}{(2 \pi r_0 \Delta )^3} 
\exp(-\frac{{\bf p}^2}{2 \Delta ^2} - \frac{{\bf r}^2}{2r_0^2}) 
\delta (t). 
\label{8.71} 
\end{equation} 
 
Note that this ansatz corresponds to 
the independent one-particle sources of Kopylov and Podgoretsky, 
all of the same type, 
characterized by a universal size of $\sim (2\Delta)^{-1}$, 
with the centers 
distributed according to a Gaussian of a dispersion 
$\widetilde{r}_0^2=r_0^2-(2\Delta)^{-2}$ (see Eq. (\ref{21A})). 
Then, in the low density limit 
but regardless of the validity of the {\it smoothness} condition 
$\widetilde{r}_0 \gg (2\Delta)^{-1}$ (see, however, the footnote after 
Eq. (\ref{8.10}) concerning the independence assumption), 
the correlation function of two non-interacting identical 
particles measures the 
dispersion of the relative 4-coordinates $\widetilde{x}$ of the 
centers of the one-particle sources 
as the inverse width squared of the correlation effect seen in the 
relative momenta ${\bf q}={\bf p}_1-{\bf p}_2$ 
\cite{pod89}. For spin-0 bosons 
\begin{equation} 
R({\bf p}_1,{\bf p}_2)=1+\langle \cos(q\widetilde{x})\rangle = 
1+\exp(-\widetilde{r}_0^2{\bf q}^2). 
\label{RKP} 
\end{equation} 
 
In this model the initial boson phase space density 
(not affected by the BE effect) is given by 
\begin{equation} 
\tilde{f}_n({\bf p},{\bf x})=\frac{n}{(r_0 \Delta)^3} 
\exp(- \frac{{\bf p}^2}{2\Delta^2} 
- \frac{{\bf x}^2}{2r_0^2}). 
\label{8.31} 
\end{equation} 
The mean densities 
at a fixed boson momentum {\bf p} and 
averaged over all phase space 
are 
\begin{equation} 
\langle \tilde{f}_n \rangle_{{\bf p}} \equiv 
\int d^3{\bf x}(\tilde{f}_n)^2~\left/\right. 
\int d^3{\bf x}\tilde{f}_n = 
\frac{n}{(\sqrt{2}r_0 \Delta)^3} 
\exp(- \frac{{\bf p}^2}{2\Delta^2}) 
\label{8.51} 
\end{equation} 
and 
\begin{equation} 
\langle \tilde{f}_n \rangle \equiv 
\int d^3{\bf x}d^3{\bf p}(\tilde{f}_n)^2~\left/\right. 
\int d^3{\bf x}d^3{\bf p}\tilde{f}_n = 
n\left/\right.(2 r_0 \Delta)^3, 
\label{8.41} 
\end{equation} 
respectively. 
Similarly, the initial inclusive densities $\tilde{f}({\bf p},{\bf x})$, 
$\langle \tilde{f} \rangle_{{\bf p}}$ and $\langle \tilde{f} \rangle$ 
are given by 
Eqs. (\ref{8.31})-(\ref{8.41}) with the 
multiplicity $n$ substituted by the initial mean multiplicity. 
 
It is worth noting an approximate equality 
(see also a model independent prove in \cite{ber94}) 
between the mean phase space 
density in the low density limit 
\begin{equation} 
\langle \tilde{f}\rangle_{{\bf p}} = 
\frac{\eta}{(\sqrt{2}r_0 \Delta)^3} 
\exp(- \frac{{\bf p}^2}{2\Delta^2}) 
\doteq \frac{\pi^{3/2}}{r_0^3}N({\bf p}) 
\label{8.51a} 
\end{equation} 
and the mean number of pions building the BE 
enhancement in the two-pion correlation function 
\begin{equation} 
\int d^3{\bf q}[R(p_1,p_2)-1]N({\bf p}_1)N({\bf p}_2)/N({\bf p})= 
\frac{\pi^{3/2}}{\widetilde{r}_0^3}N({\bf p})\approx 
\langle \tilde{f}\rangle_{{\bf p}}, 
\label{8.51aa} 
\end{equation} 
${\bf p}_{1,2}={\bf p}\pm {\bf q}/2$. 
This equality is valid up to relative corrections 
$O((r_0\Delta)^{-2})$ and $O(\langle \tilde{f}\rangle_{{\bf p}})$, 
the latter representing an impact of the BE correlations on the 
single-boson spectrum (see Section \ref{sec5b}). 
 
It is important that the Gaussian ansatz in Eq. (\ref{8.71}) 
allows to express the functions 
$G_n({\bf p}_1,{\bf p}_2)$ and the integrals $g_n$ 
(see Eqs. (\ref{8.9})) 
in simple analytical forms \cite{alinote95}: 
\begin{eqnarray} 
G_n({\bf p}_1,{\bf p}_2) &=& (2 \pi \Delta^2A_n  )^{-3/2} 
\exp(-b_n^+({\bf p}_1+{\bf p}_2)^2- b_n^-({\bf p}_1-{\bf p}_2)^2), 
\nonumber \\ 
g_n&=&(8 \Delta^2A_n b_n^+)^{-3/2}, 
\label{8.72} 
\end{eqnarray} 
where $A_n$, $b_n^+$ and $b_n^-$ are given by the recurrence 
relations: 
\begin{eqnarray} 
A_n&=&2\Delta^2 A_{n-1}(b_{n-1}^++b_{n-1}^-+b_1^++b_1^-), 
\nonumber \\ 
1/b_n^+&=&1/(b_{n-1}^++b_1^+)+ 1/(b_{n-1}^-+b_1^-), 
\nonumber \\ 
b_n^-&=&b_1^+b_1^-/b_n^+, 
\label{8.73} 
\end{eqnarray} 
with $A_1=1$, $b_1^+=1/(8\Delta^2)$ and $b_1^-=r_0^2/2$. 
The recurrence relations of this type allow for the analytical 
solution \cite{zim97}. In our case it reads as: 
\begin{equation} 
b_n^+ = b_1^+\epsilon^{-1}(1-\rho^n)\left/\right.(1+\rho^n),~~~ 
A_n = \beta^{2n/3}\epsilon(1-\rho^{2n}),~~~ 
g_n = \beta^{-n}(1-\rho^n)^{-3}, 
\label{exact} 
\end{equation} 
where $\epsilon^{-1}=2r_0\Delta$, $\rho=(1-\epsilon)/(1+\epsilon)$ and 
the parameter 
\begin{equation} 
\beta =(r_0\Delta+1/2)^3 
\label{beta} 
\end{equation} 
can be considered as a characteristic phase space volume. 
 
For example, for $n=2$ and 3, we have: 
$A_2=(1+\epsilon^{-2})/2$, $A_3=(1+3\epsilon^{-2})^2/16$, 
$b_2^+=2b_1^+/(1+\epsilon^2)$, 
$b_3^+=3b_1^+(1+\epsilon^2/3)/(1+3\epsilon^2)$, 
$g_2=\epsilon^3$, $g_3=[4\epsilon^2/(3+\epsilon^{2})]^3$. 
Recall that the cumulants related to $g_2$ and $g_3$ are 
(see Eq. (\ref{8.131})) $K_2=g_2$ and $K_3=2g_3$. 
 
It follows from the recurrence relations (\ref{8.73}) or their 
analytical solutions (\ref{exact}) that 
the slope parameters 
$b_n^+$ and $b_n^-$ approach each other with increasing $n$. 
In the large-$n$ ($n > r_0\Delta$) limit we then have \cite{alinote95}: 
\begin{equation} 
b_n^+\rightarrow b_n^-\rightarrow r_0/(4\Delta),~~~ 
A_n\rightarrow \beta^{2n/3}\left/\right.(2r_0\Delta),~~~ 
g_n\rightarrow \beta^{-n} 
\label{n>} 
\end{equation} 
and 
\begin{equation} 
G_n({\bf p}_1,{\bf p}_2) \rightarrow 
\beta^{-n}\left(\frac{r_0}{\pi\Delta}\right)^{3/2} 
\exp\left(-\frac{r_0}{4\Delta}(4{\bf p}^2+{\bf q}^2)\right). 
\label{Gn>} 
\end{equation}

In very large-$n$ ($n>e\beta$) limit, 
using the large-$n$ behavior of the parameters $g_n$, we can 
get from the recurrence relation (\ref{8.10}) 
the following behavior of the BE weight \cite{alinote95}: 
\begin{equation} 
\omega_n \rightarrow c(\beta)n!\left/\right.\beta^n, 
\label{8.10a} 
\end{equation} 
where $c(\beta)$ is a 
function factorially increasing with $\beta$, $c(1)=1$.\footnote 
{A good approximation is 
$c(\beta)\doteq\beta^{d(\beta)}$, $d(\beta)=a_1+a_2\beta^{a_3}$, 
$a_1=0.617$, 
$a_2=0.621$ and $a_3=0.788$.} 
 
It is worth noting that the large-$n$ limits become equalities at 
$\beta=1$ ($r_0\Delta=1/2$) when $g_n=A_n=1$, $K_n=(n-1)!$, $\omega_n=n!$ 
and $b_n^-=b_n^+=r_0/(4\Delta)$. 
Recall that $\beta=1$ corresponds to the minimal 
possible phase space volume when all the particle emitters are 
situated at one and the same space-time point so that the size 
$(2\Delta)^{-1}$ of the 
elementary source determines not only the width of the single-particle 
spectrum but also the characteristic distance between the production 
points (see however the footnote after Eq. (\ref{8.10}) and also 
ref. \cite{ame-led} for a more detailed discussion). 
In such a case $\widetilde{r}_0=0$ and the correlation function 
equals 2 for any value of ${\bf q}$. 
 
In the low-$n$ ($n < r_0\Delta$) limit, {\it i.e.} in the case of a 
large phase space volume, it follows from Eqs. 
(\ref{8.73}) or (\ref{exact}) that the slope parameter 
$b_n^+$ increases linearly with $n$ up to the corrections 
$O((2r_0\Delta)^{-2})$ 
and that, at $n>2r_0\Delta$, this increase 
saturates at $r_0/(4\Delta)$. 
Similar behavior shows the parameter $A_n/\beta^{2n/3}$. Thus, 
at $n\ll 2r_0\Delta$, we have: 
\begin{equation} 
b_n^+=b_1^+b_1^-/b_n^-\doteq nb_1^+,~~~ 
A_n\doteq n(r_0\Delta)^{2(n-1)},~~~ 
g_n\doteq n^{-3}(r_0\Delta)^{-3(n-1)}. 
\label{n<} 
\end{equation} 
Comparing the low-$n$ approximations (\ref{n<}) for the parameters 
$b_n^+$, $A_n$ and $g_n$ with the large-$n$ ones in Eqs. (\ref{n>}), 
we can see that they tail each other 
at $n=2r_0\Delta$, $r_0\Delta/2$ and $r_0\Delta$ respectively. 
Correspondingly, the low-$n$ approximation for the 
$G_n$-function 
\begin{equation} 
G_n({\bf p}_1,{\bf p}_2) \doteq r_0{}^3(r_0\Delta)^{-3n}(2\pi n)^{-3/2} 
\exp(-{\bf p}^2n/2\Delta^2- {\bf q}^2r_0{}^2/2n) 
\label{Gn<} 
\end{equation} 
tails with the large-$n$ one in Eq. (\ref{Gn>}) at $n=n_t$, 
$1/2<n_t/(r_0\Delta)<2$. 
 
Consider now the correlation function $R_n$ defined in Eq. (\ref{cf2}). 
To determine the normalization constant $c_n$, it is convenient to 
rewrite the single- and two-boson spectra at a fixed multiplicity $n$ 
as 
\begin{equation} 
\begin{array}{c} 
N_n^{(1)}({\bf p})= 
\sum_{j=0}^{n-1}\frac{\omega_{n-1-j}\left/\right.(n-1-j)!}{\omega_n/n!} 
G_{j+1}({\bf p},{\bf p})\equiv 
\sum_{j=0}^{n-1}\frac{w(n-1-j)}{w(n)}\widetilde{G}_{j+1}({\bf p},{\bf p}), 
\\ \\ 
{N}_n^{(2)}({\bf p}_1,{\bf p}_2) = 
\sum_{j=0}^{n-2} 
\frac{\omega_{n-2-j}/(n-2-j)!}{\omega_n/n!} 
\sum_{l=0}^j[G_{l+1}({\bf p}_1,{\bf p}_1) 
G_{j-l+1}({\bf p}_2,{\bf p}_2)+~~~~~~~~~~~~~~~ 
\\ 
~~~~~~~~~~~~~~~~~~~~~~~~~~~~~~~~~~~~~G_{l+1}({\bf p}_1,{\bf p}_2) 
G_{j-l+1}({\bf p}_2,{\bf p}_1)] 
\\ \\ 
\equiv 
\sum_{j=0}^{n-2} 
\frac{w(n-2-j)}{w(n)} 
\sum_{l=0}^j[\widetilde{G}_{l+1}({\bf p}_1,{\bf p}_1) 
\widetilde{G}_{j-l+1}({\bf p}_2,{\bf p}_2)+ 
\widetilde{G}_{l+1}({\bf p}_1,{\bf p}_2) 
\widetilde{G}_{j-l+1}({\bf p}_2,{\bf p}_1)], 
\label{n_n} 
\end{array} 
\end{equation} 
where $w(n)$, defined in Eq. (\ref{mult1}), 
coincides with the BE affected multiplicity distribution arising from 
the Poissonian one characterized by the initial mean multiplicity $\eta$ 
and 
$\widetilde{G}_{i}({\bf p}_1,{\bf p}_2)=\eta^i 
G_i({\bf p}_1,{\bf p}_2)$. Noting further that $b_n^+$ approaches 
the limiting value $r_0/(4\Delta)$ from below, while $b_n^-$ 
does it from above, we can see from Eq. (\ref{8.72}) that, at large $q$, 
all terms in Eqs. (\ref{n_n}) for ${N}_n^{(1)}({\bf p}_{1,2})$ and 
${N}_n^{(2)}({\bf p}_1,{\bf p}_2)$ 
(${\bf p}_{1,2}={\bf p}\pm {\bf q}/2$) 
can be neglected except for those containing the lowest slope $b_1^+$. 
For the normalization constant 
$c_n=\lim_{q\to\infty}[{N}_n^{(1)}({\bf p}_{1}) 
{N}_n^{(1)}({\bf p}_{2})\left/\right.{N}_n^{(2)}({\bf p}_1,{\bf p}_2)]$ 
in Eq. (\ref{cf2}) for the correlation function 
we thus get \cite{alinote95}: 
\begin{equation} 
c_n= [w(n-1)]^2\left/\right.[w(n)w(n-2)]\equiv 
n\omega_{n-1}^2\left/\right.[(n-1)\omega_n\omega_{n-2}]. 
\label{cn} 
\end{equation} 
 
Note that $c_2=2/\omega_2 \in (1,2)$; with the increasing multiplicity 
$c_n$ decreases and, according to Eq. (\ref{8.10a}), $c_n\doteq 1$ for 
$n > e\beta$. For large phase space volumes (when $\omega_n\doteq 1$ 
at small $n$), the normalization $c_n\doteq n/(n-1)$ and the exclusive 
correlation function $R_n$, normalized to 1 at large $|{\bf q}|$, 
becomes close to the usual definition as a ratio of the two-particle 
spectrum to the product of the single-particle ones, both 
spectra normalized to 1. Generally, the latter definition is 
however not reliable since it leads to the plateau height 
of $\omega_n\omega_{n-2}/\omega_{n-1}^2 > 1$ which, 
in case of a small phase space volume $\beta$ and a small $n$, 
can be substantially larger than 1. 
For example, for $n=2$ this height is $\omega_2$ and can reach a value 
of 2 if $\beta\rightarrow 1$. 
 
Regarding the (semi-)inclusive single- and two-boson spectra, 
they can be written in a form similar to Eqs. (\ref{n_n}) only 
in the initially Poissonian case: 
\begin{eqnarray} 
{N}^{(1)}({\bf p})&=&\sum_{n} 
\sum_{j=0}^{n-1}w(n-1-j)\widetilde{G}_{j+1}({\bf p},{\bf p})\left/\right. 
\sum_{n} w(n), 
\nonumber \\ 
{N}^{(2)}({\bf p}_1,{\bf p}_2) &=& \sum_{n} \sum_{j=0}^{n-2}w(n-2-j) 
\sum_{l=0}^j[\widetilde{G}_{l+1}({\bf p}_1,{\bf p}_1) 
\widetilde{G}_{j-l+1}({\bf p}_2,{\bf p}_2)+~~~~~~~~~~~~~~~~~~ 
\nonumber \\ 
&&~~~~~~~~~~~~~~~~~~~~~~~~~~~~~~\widetilde{G}_{l+1}({\bf p}_1,{\bf p}_2) 
\widetilde{G}_{j-l+1}({\bf p}_2,{\bf p}_1)]\left/\right.\sum_{n} w(n). 
\label{n_n1n2} 
\end{eqnarray} 
The normalization constant in Eq. (\ref{cf2i}) for the (semi-)inclusive 
correlation function is then \cite{alinote95}: 
\begin{equation} 
c= [\sum_{n} w(n-1)]^2\left/\right.[\sum_{n} w(n)\sum_{n} w(n-2)]. 
\label{c} 
\end{equation} 
Clearly, in the completely inclusive case 
(when the sums include all $n$ from 0 to $\infty$ 
and $\sum_{n} w(n-j)=1$), we have $c=1$ and 
\begin{eqnarray} 
\langle n\rangle &=&\sum_{j=0}^{\infty} 
\widetilde{g}_{j+1} \equiv \widetilde{g}, 
\nonumber \\ 
{N}^{(1)}({\bf p})&=&\sum_{j=0}^{\infty} 
\widetilde{G}_{j+1}({\bf p},{\bf p}) \equiv \widetilde{G}({\bf p},{\bf p}), 
\nonumber \\ 
{N}^{(2)}({\bf p}_1,{\bf p}_2) &=& \sum_{j=0}^{\infty} 
\sum_{l=0}^j[\widetilde{G}_{l+1}({\bf p}_1,{\bf p}_1) 
\widetilde{G}_{j-l+1}({\bf p}_2,{\bf p}_2)+ 
\widetilde{G}_{l+1}({\bf p}_1,{\bf p}_2) 
\widetilde{G}_{j-l+1}({\bf p}_2,{\bf p}_1)] 
\nonumber \\ 
&\equiv & 
\widetilde{G}({\bf p}_1,{\bf p}_1)\widetilde{G}({\bf p}_2,{\bf p}_2)+ 
\widetilde{G}({\bf p}_1,{\bf p}_2)\widetilde{G}({\bf p}_2,{\bf p}_1), 
\label{n_inc} 
\end{eqnarray} 
where $\widetilde{g}_n=\eta^n g_n$. For the inclusive 
correlation function (\ref{cf2i}) 
we have \cite{alinote95}: 
\begin{equation} 
R({\bf p}_1,{\bf p}_2)= 1+\frac{ 
\widetilde{G}({\bf p}_1,{\bf p}_2)\widetilde{G}({\bf p}_2,{\bf p}_1)} 
{\widetilde{G}({\bf p}_1,{\bf p}_1)\widetilde{G}({\bf p}_2,{\bf p}_2)}. 
\label{cfinc} 
\end{equation} 
Thus, in the considered case of the initially 
Poissonian multiplicity distribution, 
the intercept $R({\bf p},{\bf p})\equiv R(0)=2$ 
in agreement with the result generally valid for 
thermalized systems \cite{sl94}. 
Note that Eq. (\ref{cfinc}) coincides with similar expressions 
in refs. \cite{cha95,zim97} up to a normalization factor 
$\langle n\rangle^2/\langle n(n-1)\rangle$. 
With the increasing density, the latter decreases from 1 to 1/2 
and, for dense systems, forces the corresponding correlation 
function to 1. Such a behavior was incorrectly 
interpreted \cite{cha95,zim97} as a coherent effect 
(see also a discussion in Section \ref{sec6}).

\section{Results}\label{sec5} 
\subsection{Multiplicity distributions}\label{sec5a} 
We will consider here the multiplicity distribution (\ref{mult1}) 
resulting due to the BE effect on the initially 
Poissonian one with the mean multiplicity $\eta$. 
In accordance with the large-$n$ behavior of the BE weights in 
Eq. (\ref{8.10a}), it takes on the following limiting form at 
$n> e\beta$ \cite{alinote95}: 
\begin{equation} 
w(n)\rightarrow {\rm const'}\cdot \xi^n,~~~~ 
\xi=\eta/\beta . 
\label{mult} 
\end{equation} 
 
The large-$n$ behavior of the multiplicity distribution in Eq. (\ref{mult}) 
indicates that it approaches the BE one: 
\begin{equation} 
w_{{\rm BE}}(n)=\nu^n\left/\right.(1+\nu)^{n+1},~~~~ 
\nu=\xi/(1-\xi), 
\label{multBE} 
\end{equation} 
with the mean multiplicity $\nu$.\footnote 
{ 
This is in accordance with the appearance of the Bose-condensate in 
a dense ideal Bose gas \cite{landau}. 
The fluctuations of the number of particles in the condensate are 
very large - they are described by the well-known Einstein formula 
for identical bosons in the same quantum state. 
The corresponding BE multiplicity distribution in Eq. (\ref{multBE}) turns to 
the Reley one for very large mean multiplicities. 
This type of BE condensate should not be mixed up with the multiboson 
coherent (laser) state in which the BE correlations are absent 
and the multiplicity distribution corresponds to the Poisson law. 
} 
This is demonstrated in Figs. 1 and 2. Thus, at $r_0=2.1$ fm 
and $\Delta=0.25$ GeV/c, the BE effect transforms 
the initial Poissonian multiplicity distribution with $\eta=30$ 
(dotted curve in Fig. 1a) 
to the one with much higher mean and 
dispersion values (solid curve in Fig. 1a). 
The exponential tail expected for the BE distribution 
is clearly seen in Fig. 1b where the 
results are presented in logarithmic scale for $\eta=10$, 
$\Delta=0.25$ GeV/c and $r_0=1.5$ fm. 
One may see that Eq. (\ref{mult}) (dashed line) becomes 
an excellent approximation
for $n>30$, which is close to the condition $n>e\beta=37.6$ for the present
choice of parameters. 
The slope parameter $b$ in the exponential fit 
$ w(n) = {\rm const}\cdot \exp(-bn)$ 
of this tail at large $n$ should be, 
according to Eq. (\ref{mult}), 
only a function of the variable $\xi$: $b=-\ln(\xi)$. 
Such a scaling is demonstrated in Fig. 2a for various values 
of $\eta$, $\Delta$ and $r_0$. 
Note that $\xi =$ 0.95 and 0.72 for Figs. 1a and 1b, 
corresponding to $b=$ 0.02 and 0.27, respectively.\footnote 
{ 
At the explosion point $\xi=1$ the tail of the multiplicity 
distribution becomes a constant ($b=0$) so that the mean multiplicity 
$\langle n\rangle$ would go to infinity 
provided that there are no energy-momentum constraints. 
Note that the corresponding critical initial mean multiplicity 
$\eta_{cr}=\beta\equiv (r_0\Delta+1/2)^3$ is close but different from that 
given in Eq. (9) of ref. \cite{pra93}. 
For the origin of this difference see discussion in Section \ref{sec6}. 
} 
 
It should be noted that the experimental data point to a moderate 
value of the density parameter $\xi$. Thus, 
taking 0.2 as an estimate of the inclusive phase space density 
at ${\bf p}=0$ 
from AGS and SPS experiments and using Eq. (\ref{8.51a}), we get 
(see the last Section) 
$\xi\approx 0.4-0.5$. 
 
Using Eq. (\ref{n_inc}) for $\langle n\rangle$ 
and tailing the large- and small-$n$ approximations of the integrals 
$g_n$ at $n_t=r_0\Delta$ (see 
Eqs. (\ref{n>}) and (\ref{n<})), we can approximate the mean 
multiplicity as 
\begin{eqnarray} 
\langle n\rangle &\doteq & 
\left[1+\tilde{\xi}/2^{3}+\cdots+ 
\tilde{\xi}^{n_t-1}/n_t{}^{3}\right]\eta 
+\xi^{n_t}\nu 
\nonumber \\ 
&\equiv& \langle n\rangle_{{\rm g}}+\langle n\rangle_{{\rm c}}, 
\label{n-appr} 
\end{eqnarray} 
where $\tilde{\xi}=\eta/(r_0\Delta)^3 > \xi$; 
the density parameters $\tilde{\xi}$ and $\xi$ coincide at 
$r_0\Delta\gg 1$. 
At large phase space volumes, $(r_0\Delta)^3\gg 1$, 
the two terms in Eq. (\ref{n-appr}) can be considered as 
contributions of the BE gas and BE condensate respectively. 
It can be seen that the condensate dominates on condition 
$\langle n\rangle > \beta$. 
 
Note that in the rare gas limit $\tilde{\xi}\ll 1$, 
we have $\langle n\rangle \doteq \left[1+\tilde{\xi}/2^{3}\right]\eta 
\equiv \eta+K_2\eta^2$, {\it i.e.} the increase of the 
mean multiplicity is dominated by the contribution of the 
second order cumulant $K_2=(2r_0\Delta)^{-3}$. 
The corresponding multiplicity distribution then becomes 
somewhat wider than the Poissonian one 
(see Eqs. (\ref{oappr}), (\ref{mult1})): 
$w(n)= {\rm const}\cdot (1+C_2^nK_2)\eta^n/n!$. 
 
In Fig. 2c we demonstrate the approach of the mean multiplicity 
$\langle n\rangle$ to the limiting scaling value 
$\nu=\xi/(1-\xi)$, 
though only for $\xi$ very close to the explosion point $\xi=1$ 
($\xi >0.99$). Instead, in the region of $\xi<0.9$ 
indicated by present experiments, 
we can see, in agreement with Eq. (\ref{n-appr}) 
an approximate $\xi$-scaling of the ratio 
$\langle n\rangle/\eta$ (Fig. 2b).

Since, in the realistic event generators, the multiboson effects can be 
accounted for only in the lowest order cumulant approximation 
\cite{ame-led}, it is instructive to study the saturation of these 
effects with the increasing number $N_{cum}$ of the contributing 
cumulants. In Fig. \ref{Ncum} we show $N_{cum}$-dependence of the 
ratio $\langle n\rangle/\eta$ of the BE affected mean multiplicity 
to the initial one for different values of the density parameter $\xi$. 
For example, at $\xi=0.8$ this ratio saturates at $N_{cum}\sim 10$ 
($\sim 40\%$ increase of $\langle n\rangle$). 
At $N_{cum}=4$, representing a practical limit due to the numerical problems 
\cite{ame-led}, the effect is underestimated by $\sim 25\%$ 
($\langle n\rangle/\eta\approx$ 1.3 instead of 1.4). 
The situation is more optimistic for lower densities. Thus 
at $\xi=0.5$ the effect ($\sim 15\%$ increase of $\langle n\rangle$) 
practically saturates at $N_{cum}=4$.

Up to now, we have considered the symmetrization effect on the production of 
only one type of identical bosons. For a system of charged and neutral pions,
the total symmetrization weight in the model coincides with a product
of the separate BE weights $\omega_{n_+}$, $\omega_{n_-}$ and $\omega_{n_0}$
provided the pions are emitted in  unpolarized and uncorrelated isospin states,
the effect of their FSI is negligible and there are practically 
no restrictions due to
energy--momentum and isospin constraints. The latter assumption may be
reasonable at high energies when the
subsystem of interfering pions represents a small part of the produced
multiparticle system.
The initial distribution of pion species is then a trinomial one,
strongly peaked at $n_+=n_-=n_0=n/3$:
$w(n_+,n_-,n_0)=n!/(3^n n_+!n_-!n_0!)$.
After the symmetrization in the large--$n$ condensate  limit 
($\omega_{n_i}\doteq n_i!/\beta^{n_i}$), it
becomes independent of $n_i$ and yields a substantial probability
($2(n-n_0+1)/[(n+1)(n+2)]$) of any value of $n_0$ at a fixed total
pion multiplicity $n$. Particularly, 
the production of
so called Centauro (anti--Centauro) events containing mainly  charged 
(neutral) pions then becomes possible.
The probability of the extreme charge configurations can be 
enhanced even stronger
in case of isospin constraints, for example, if pions were
produced in isosinglet pairs \cite{pra93,pra94,kk86}.
The latter mechanism can be of particular importance in 
the near--threshold multipion production,\footnote{ 
A proposal of an experimental study of the near--threshold multipion 
system at Serpukhov accelerator was recently discussed by 
V.A.~Nikitin. Similar idea was also communicated to us by 
L.L.~Nemenov.} 
due to the limited total isospin and charge. Unfortunately, the 
$n!$ enhancement of the near--threshold condensate production 
will be more than 
compensated by the phase space suppression factor of 
$(\bar{p}_n/\Delta)^{3n}\sim (n!)^{-3/2}$, where 
$\bar{p}_n=[2m(\sqrt{s}-\sum_i m_i)/n]^{1/2}$ is the mean pion momentum 
near threshold.   
 
\subsection{Single-particle spectra}\label{sec5b} 
The influence of the BE effect on the single-boson 
spectrum for a given boson multiplicity $n$, 
can be seen from Eqs. (\ref{8.131})-(\ref{8.12}), 
(\ref{cf1}) and (\ref{8.72}). 
At sufficiently large momenta, when the 
local density $\langle f_n\rangle_{{\bf p}}$ 
remains small even at large $n$, this spectrum is dominated by 
the contribution $\beta_n P({\bf p})$, 
$\beta_n =n\omega_{n-1}/\omega_n$, 
of the initial spectrum. 
In such a rare gas limit, 
$\xi_{n,{\bf p}}\equiv \xi_n\exp(-{\bf p}^2/2\Delta^2)\ll 1$, 
we can write (see Eqs. (\ref{8.131}), (\ref{cf1}) and (\ref{oappr})): 
\begin{eqnarray} 
{N}_n({\bf p}) &\doteq& n\frac{\omega_{n-1}}{\omega_{n}}G_1({\bf p},{\bf p}) 
+n(n-1)\frac{\omega_{n-2}}{\omega_{n}}G_2({\bf p},{\bf p}) 
\nonumber \\ 
&\doteq& n[1-(n-1)K_2]G_1({\bf p},{\bf p})+n(n-1)G_2({\bf p},{\bf p}) 
\nonumber \\ 
&\doteq& nP({\bf p})+n(n-1)K_2[2^{3/2}P(2^{1/2}{\bf p})-P({\bf p})]. 
\label{Pappr} 
\end{eqnarray} 
Otherwise, at large local densities, ${N}_n({\bf p})$ 
is determined by the asymptotic large-density spectrum \cite{alinote95}: 
\begin{equation} 
{N}_n({\bf p}) \rightarrow 
n\left(\frac{r_0}{\pi\Delta}\right)^{3/2} 
\exp\left(-\frac{r_0}{\Delta}{\bf p}^2\right) 
\equiv nP_{{\rm c}}({\bf p}), 
\label{Pasy} 
\end{equation} 
associated with the BE condensate and 
corresponding to the asymptotic (large-$n$) value 
$r_0/(4\Delta)$ 
of the slope parameters $b_n^{\pm}$. 
Note that $P_{{\rm c}}({\bf p})$ is normalized to unity and that, 
at $\beta=1$, 
it coincides with the initial distribution $P({\bf p})$. 
 
It is clear from Eqs. (\ref{Gn>}), (\ref{8.10a}) and (\ref{n_n}) that for small momenta, 
$p < \Delta(r_0\Delta-1/2)^{-1/2}$, 
the condensate regime in Eq. (\ref{Pasy}) settles on condition 
$\xi_n > e$. 
For larger momenta, we must take into account that the condensate 
contribution vanishes much faster than that of BE gas, thus leading 
to much stronger condition of the condensate dominance: 
\begin{equation} 
\xi_n > (2r_0\Delta)^{-3/2}\exp[(2r_0\Delta-1)p^2/2\Delta^2]. 
\label{concon} 
\end{equation}

Similar to the fixed multiplicity case, the inclusive single-boson spectrum 
at small local densities 
tends to $\eta P({\bf p})\equiv \widetilde{G}_{1}({\bf p},{\bf p})$ 
and, at large ones, it approaches 
the asymptotic high-density spectrum 
(see Eq. (\ref{Pasy})): 
\begin{equation} 
{N}({\bf p}) \rightarrow 
\langle n\rangle P_{{\rm c}}({\bf p}). 
\label{Pasyinc} 
\end{equation} 
The transfer of the initial spectrum to the high-density one is demonstrated 
for the inclusive distribution 
in Fig. 4a and, more clearly, 
for $\xi$ closer to the explosion point $\xi=1$, in Fig. 4b.\footnote 
{These results agree with those obtained in refs. 
\cite{pra93,pra94} (see also \cite{sl94,ame-led} and references therein) 
except for an incorrect conclusion \cite{pra93} that the width 
of the narrow peak due to the BE "condensate" 
is of $1/r_0$.} 
 
Tailing the large- and small-$n$ behavior of the $G_n$-functions 
at $n_t\sim r_0\Delta$ (see 
Eqs. (\ref{Gn>}) and (\ref{Gn<})), we can approximate the inclusive 
single-boson spectrum in Eq. (\ref{n_inc}) as 
\begin{eqnarray} 
{N}({\bf p}) &\doteq & 
\left[1+\tilde{\xi}_{{\bf p}}\left/\right.2^{3/2}+\cdots+ 
\tilde{\xi}_{{\bf p}}{}^{n_t-1}\left/\right.n_t{}^{3/2}\right]\eta P({\bf p}) 
+\xi^{n_t}\nu P_{{\rm c}}({\bf p}) 
\nonumber \\ 
&\equiv& {N}_{{\rm g}}({\bf p})+{N}_{{\rm c}}({\bf p}), 
\label{Pasyincpar} 
\end{eqnarray} 
where $\tilde{\xi}_{{\bf p}}=(2\pi)^{3/2}\eta P({\bf p})/r_0{}^3\equiv 
\tilde{\xi}{\rm exp}(-{\bf p}^2\left/\right.2\Delta^2)$. 
Clearly, for large phase space volumes, $(r_0\Delta)^3\gg 1$, 
the two terms in Eq. (\ref{Pasyincpar}) can be interpreted as 
contributions of the BE gas and BE condensate respectively. 
Like in the fixed multiplicity case, 
the condensate dominates on condition (\ref{concon}) with the 
substitution $n \rightarrow \langle n\rangle$ 
($\xi_n \rightarrow \langle n\rangle/\beta$). 
 
In Fig. 4 we compare 
the inclusive single-boson spectra with the approximate formula 
(\ref{Pasyincpar}). A good agreement is obtained 
despite the calculations were done for not very 
large phase space volumes. 
Some underestimation of $N({\bf p})$ at intermediate local densities
$\widetilde{\xi}_{{\bf p}}$ (Fig. 4a) 
and the corresponding underestimation of
$\langle n\rangle$ at moderate densities $\xi$ (Fig. 2b) become weaker
for larger systems (larger multiplicities in Fig. 2b) due to increasing
number $n_t\sim r_0\Delta$ of the terms in Eqs. (\ref{n-appr}), 
(\ref{Pasyincpar}) and thus -
decreasing relative contribution of the tailing region. We may conclude
that the accuracy of Eqs. (\ref{n-appr}) and (\ref{Pasyincpar}) 
is reasonable for the systems
produced in heavy ion collisions at SPS and that it will be even better
for larger systems at RHIC and LHC. 
Experimentally the effect of BE "condensate" 
was searched for at SPS CERN as a low-$p_t$ 
enhancement, however, with rather uncertain results 
%(see, {\it e.g.}, \cite{qm,quarkmatter96}). 
(see, {\it e.g.}, \cite{qm}). 
 
It follows from Eq. (\ref{Pasyincpar}) that for sufficiently large 
($n_t=r_0\Delta \gg 1$) and not very dense ($\xi \ll 1$) systems, 
similarly to the $\xi$-scaling of $\langle n\rangle/\eta$ (see Fig. 2b), 
the ratio ${N}({\bf p})/[\eta P({\bf p})]$ scales with the local density 
parameter $\tilde{\xi}_{{\bf p}}$. 
It appears that analogical scaling takes place also at fixed multiplicity 
$n$.\footnote 
{In this case, 
due to the explicit dependence of the particle spectra 
on the complicated BE weights $\omega_n$, 
there is no analytical approximation similar to Eq. (\ref{Pasyincpar}).} 
In Fig. 5 we show the ratio of the 
single-particle spectrum at fixed $n$ to the dominant 
large-${\bf p}$ contribution 
$\beta_n P({\bf p})$ of the initial spectrum calculated at ${\bf p}=0$ 
as a function of $\xi_n$ for various multiplicities $n$. 
An approximate $\xi_n$-scaling is seen up to 
$\xi_n$ of the order of unity. At larger $\xi_n$ this ratio 
approaches the condensate limit $(2r_0\Delta)^{3/2}\xi_n$ which 
no more scales with $\xi_n$ 
(see the corresponding curves in Fig. \ref{ratsp}b). 
What scales at large $\xi_n$ is not the ratio 
of the two contributions but the ratio of their integrals, $n/\beta_n$, 
the limiting value of which is just equal to $\xi_n=n/\beta$ 
since, according to Eq. (\ref{8.10a}), 
$\beta_n =n\omega_{n-1}/\omega_n \rightarrow \beta$ for 
$n>{\rm e}\beta$. 

Note that, in the absence of simple analytical approximations for the
fixed-$n$ spectra, the approximate $\xi_n$- or $\xi_{n,{\bf p}}$--scaling
can be used to overcome technical problems with factorially large numbers
at high multiplicities. For example, $N_n(0)$ at a given density $\xi_n<1$
can be obtained by calculating $N_{n'}(0)$ at a smaller multiplicity
$n'<n$ keeping the same density $\xi_{n'}=\xi_n$ and then rescaling to
$N_n(0)\doteq (\beta_n/\beta_{n'})N_{n'}(0)$, where
$\beta_n/\beta_{n'}\doteq 1$ for $n' > e\beta$.

\subsection{Correlation functions}\label{sec5c} 
 
It follows from Eqs. (\ref{cf2}) and (\ref{8.72})-(\ref{cn}) 
that, for a given multiplicity $n$, the  correlation function intercept 
$R_n({\bf p},{\bf p})\equiv R_n(0)$ 
decreases and the correlation function width increases 
with the increasing $n$ or decreasing momentum $p$, 
both corresponding to the increasing local density parameter 
$\xi_{n,{\bf p}}$. 
 
In fact, for large local densities (see Eq. (\ref{concon})), 
the condensate behavior is achieved (see the curves in Fig. 5b) 
and the correlation function tends to unity not only at large but also 
at small ${\bf q}^2$. 
Indeed, in this limit the normalization constant 
$c_n\rightarrow 1$ (see Eqs. (\ref{c_n}) and (\ref{8.10a})) and the nominator and 
denominator of the correlation function (\ref{cf2}) 
consist of about the same 
number $\sim n^2$ of the condensate terms 
$(r_0/\pi\Delta)^3\exp[-(4{\bf p}^2+{\bf q}^2)r_0/2\Delta]$ 
(see Eqs. (\ref{Gn>}), (\ref{8.10a}) and (\ref{n_n})).\footnote 
{Of course, the absence of the correlation in the condensate limit at 
fixed multiplicity has nothing to do with the coherence effect which is 
absent in the considered model. See the footnote after Eq. (\ref{8.10}) and the 
discussion of the inclusive correlation function which appears to be 
different from 1 at whatever high densities. 
} 
The well known \cite{zaj87,zha93,pra93} lowering and widening of the 
correlation function with the increasing multiplicity is demonstrated, 
in the considered model, in Fig. 6.

Note however that in the rare gas limit, $\xi_{n,{\bf p}}\ll 1$, 
the change of the form of the correlation function with the 
increasing density is rather weak. Thus, writing in this limit 
\begin{eqnarray} 
{N}_n^{(2)}({\bf p}_1,{\bf p}_2) \doteq 
n(n-1)\frac{\omega_{n-2}}{\omega_{n}} 
[G_1({\bf p}_1,{\bf p}_1)G_1({\bf p}_2,{\bf p}_2)+ 
G_1({\bf p}_1,{\bf p}_2)G_1({\bf p}_2,{\bf p}_1)] 
\nonumber \\ 
+n(n-1)(n-2)\frac{\omega_{n-3}}{\omega_{n}} 
[G_1({\bf p}_1,{\bf p}_1)G_2({\bf p}_2,{\bf p}_2)+ 
G_1({\bf p}_1,{\bf p}_2)G_2({\bf p}_2,{\bf p}_1) 
\nonumber \\ 
+G_2({\bf p}_1,{\bf p}_1)G_1({\bf p}_2,{\bf p}_2)+ 
G_2({\bf p}_1,{\bf p}_2)G_1({\bf p}_2,{\bf p}_1)] 
\nonumber \\ 
\doteq n(n-1)[1-(2n-3)K_2]P({\bf p}_1)P({\bf p}_2) 
\left[1+\exp(-\tilde{r}_0{}^2{\bf q}^2) 
\right]~~~~~~~~~~~~~~~~~~ 
\nonumber \\ 
+n(n-1)(n-2)2^{3/2}K_2\{P({\bf p}_1)P(2^{1/2}{\bf p}_2) 
\left[1+\exp\left(-{\textstyle \frac34 } \tilde{r}_0{}^2{\bf q}^2+ 
\frac{{\bf p}{\bf q}}{2\Delta^2}\right)\right] 
\nonumber \\ 
+P({\bf p}_2)P(2^{1/2}{\bf p}_1) 
\left[1+\exp\left(-{\textstyle \frac34 } \tilde{r}_0{}^2{\bf q}^2- 
\frac{{\bf p}{\bf q}}{2\Delta^2}\right)\right]\}, 
\label{P2appr} 
\end{eqnarray} 
$\tilde{r}_0{}^2={r}_0{}^2[1-(2r_0\Delta)^{-2}]$, 
we get for the correlation function intercept an $n$-independent value 
close to 2: 
$R_n(0)\doteq 2(1-\epsilon_1)$. Here we have introduced the density 
parameter 
\begin{equation} 
\epsilon_j=j\cdot 2^{5/2}K_2\exp(-{\bf p}^2\left/\right.2\Delta^2) 
\doteq 2^{-1/2}\xi_{j,{\bf p}}. 
\label{eps} 
\end{equation} 
 
It should be noted that the correlation function $R_n({\bf q})$ 
becomes less than unity at intermediate q-values 
and approaches the limiting value of 1 from below. 
This behavior and also the related suppression of the intercept value 
is caused by the BE correlation effect on the single-particle 
spectra entering the denominator of the correlation function. 
Sometimes this distortion is corrected for by a special iterative procedure. 
Its result can be described by 
a simple low-${\bf q}^2$ correction factor:\footnote 
{The iterative correction procedure is usually used for small-acceptance 
detectors triggered by the requirement of at least two identical pions 
in the detector. The mixed reference sample then differs from the product of 
the single-particle spectra, being much more influenced by the residual 
BE correlations. The residual correlations can substantially affect also 
single-particle spectra in the case of a small effective emission volume, 
{\it e.g.}, in $e^+e^-$-collisions. 
There are also other reasons for the low-$q^2$ correction factor, 
like energy-momentum constraints or presence of dynamical correlations 
({\it e.g.}, in jets) which are destroyed in the mixed reference sample. 
For this reason the correction factor similar to that in Eq. (\ref{CFcor}) 
is often introduced as a pure phenomenological one with $A_n=1$ and $B_n$ 
treated as a free parameter. 
} 
\begin{equation} 
R_n^{cor}({\bf p}_1,{\bf p}_2) \doteq (A_n-B_n {\bf q}^2) 
R_n({\bf p}_1,{\bf p}_2). 
\label{CFcor} 
\end{equation} 
In the low density limit of our model, we have 
\begin{equation} 
A_n=1+\epsilon_{n-1},~~~~ 
B_n=\epsilon_{n-1}[1-({\bf p}\left/\right.\Delta)^2\cos^2\psi] 
\left/\right.8\Delta^2, 
\label{AB} 
\end{equation} 
where $\psi$ is the angle between the vectors ${\bf p}$ and ${\bf q}$. 
At small ${\bf q}^2$, the corrected correlation function (properly 
normalized to unity at large ${\bf q}^2$) is then 
\begin{equation} 
R_n^{cor}({\bf p}_1,{\bf p}_2) \doteq 1+\exp(-\tilde{r}_0{}^2{\bf q}^2)+ 
\epsilon_{n-2}\left[1+\exp\left(-{\textstyle \frac34 } 
\tilde{r}_0{}^2{\bf q}^2\right)\right] 
\label{Rcor} 
\end{equation} 
and can be represented in the 
usual single-Gaussian form:\footnote 
{Note that in \cite{zim97} a similar parameterization 
was used for the uncorrected 
correlation function. This led to different estimates of the 
interferometry parameters $\lambda_n$ and $r_n$ 
in the considered model.} 
\begin{equation} 
\begin{array}{c} 
R_n^{cor}({\bf p}_1,{\bf p}_2)\doteq 1+\lambda_n\exp(-r_n{}^2{\bf q}^2),\\ \\ 
\lambda_n=1+2\epsilon_{n-2},~~~~ 
r_n{}^2=\tilde{r}_0{}^2\left(1-\frac54\epsilon_{n-2}\right). 
\label{CFcorpar} 
\end{array} 
\end{equation} 
We see that with the increasing $n$ the effective interferometry 
parameters $\lambda_n$ 
and $r_n$ respectively increase and decrease slightly, starting from the 
zero density values of 1 and $\tilde{r}_0$. 
 
In the low-density limit, simple  Eqs. (\ref{Rcor}), (\ref{CFcorpar}) or directly 
Eqs. (\ref{Pappr}) and (\ref{P2appr}) allow one to determine the radius parameter $r_0$ 
by fitting the correlation functions $R_n^{cor}$ or $R_n$. 
At higher densities, however, there is no simple analytical expression for the 
correlation function $R_n$ and the eventual fit would require the use of 
rather complicated Eqs. (\ref{n_n}). Another possibility is still a simple 
single-Gaussian fit at sufficiently small q, giving the effective interferometry 
parameters $\lambda_n^{\rm eff} < 1$ and $r_n^{\rm eff} < r_0$, 
both vanishing with the increasing local density. 
The low-density radius $r_0$ and the density can 
then be determined comparing $\lambda_n^{\rm eff}$ and $r_n^{\rm eff}$ 
with the model predictions as functions of $r_0$ and $\Delta$. 
 
In Fig. 7 we show the intercept as a function of the multiplicity $n$ 
and the local density 
parameter $\xi_{n,{\bf p}}$ for several values of the mean momentum: 
$p=0$, 0.1, 0.2 and 0.4 GeV/c. 
As expected, the intercept is practically constant at low local 
densities (small $n$ or high $p$). As condensate develops, the intercept 
sharply falls down. The sharpness of this drop is however less 
pronounced at higher momenta even if plotted as a function of the 
local density parameter $\xi_{n,{\bf p}}$. 
Clearly, this lack of density scaling is related to 
a strong decrease of the condensate contribution with the increasing 
momentum. 
In fact, for the momenta $p >\Delta(r_0\Delta-1/2)^{-1/2}$ 
the low-density parameter $\xi_{n,{\bf p}}$ 
strongly overestimates the local density in the region of the condensate 
dominance (see Eq. (\ref{concon})). 
To demonstrate the possibility of the observation 
of the condensate effect for a speckle of a large number of soft pions 
not following the ordinary proportionality rule between 
the freeze-out phase space volume and pion multiplicity, 
in Fig. 7 we indicate by the arrows 
the intercept values corresponding to 
$\xi_n=3\xi=1.5$ ($n\approx 3\langle n\rangle$). 
 
In the inclusive case corresponding to the initial Poissonian 
multiplicity distribution, 
the correlation function intercept is equal to 2 for 
any local densities (see Eq. (\ref{cfinc})). 
At very large local densities 
the two-boson spectrum 
approaches twice the product of the single-boson 
ones (see Eqs. (\ref{Gn>}) and (\ref{n_inc})) 
so that the inclusive correlation function tends to the 
limiting value of 2 even at rather large relative momenta. 
The corresponding increase of the width of the 
correlation function with the increasing density parameter $\xi$ 
is demonstrated in Fig. \ref{CFI}. 
 
Note that at high local densities both the nominator and denominator of the 
correlation function at small ${\bf q}^2$ are dominated, like in the case 
of a fixed multiplicity, by the universal condensate terms, their numbers 
being about $\langle n(n-1)\rangle$ and $\langle n\rangle^2$ respectively. 
The difference between the inclusive 
and fixed multiplicity correlation functions, 
$R\rightarrow 2$ and $R_n \rightarrow 1$, is due to the fact that 
at high densities the initially Poissonian multiplicity distribution 
approaches a much wider BE one, for which 
$\langle n(n-1)\rangle =2\langle n\rangle^2$. 

Tailing the approximate equations (\ref{Gn>}) and (\ref{Gn<}) 
for the $G_n$-functions at $n_t \sim r_0\Delta$, 
we can analytically follow the behavior of the
function $\widetilde{G}({\bf p}_1,{\bf p}_2)$
(determining, according to Eq. (\ref{n_inc}),
the inclusive two-boson spectrum $N({\bf p}_1,{\bf p}_2)$) 
similar to Eq. (\ref{Pasyincpar}), modifying it by the substitutions:
$\widetilde{\xi}_{\bf p}^{n-1} \rightarrow \widetilde{\xi}_{\bf p}^{n-1}
\exp(-{\bf q}^2r_0^2/2n)$ and
$P_c({\bf p})\rightarrow P_c({\bf p})\exp(-{\bf q}^2r_0/4\Delta)$.
Particularly, for large systems ($(2r_0\Delta)^2\gg 1$), 
we have at $q\rightarrow 0$: 
\begin{equation} 
\widetilde{G}({\bf p}_1,{\bf p}_2) 
\doteq {N}_{{\rm g}}({\bf p}) \exp(-{\bf q}^2r_{{\rm g}}{}^2/2)+ 
{N}_{{\rm c}}({\bf p}) \exp(-{\bf q}^2r_{{\rm c}}{}^2/2), 
\label{Gapp} 
\end{equation} 
where $r_{{\rm c}}{}^2=r_0/(2\Delta)$ and 
\begin{equation} 
r_{{\rm g}}{}^2\doteq 
r_0{}^2\sum_{n=1}^{n_t}\tilde{\xi}_{{\bf p}}{}^n n^{-5/2}\left/\right. 
\sum_{n=1}^{n_t}\tilde{\xi}_{{\bf p}}{}^n n^{-3/2}. 
\label{rBG} 
\end{equation} 
Note that at low local densities ($\tilde{\xi}_{{\bf p}}\ll 1$) 
the effective radius $r_{{\rm g}}$ coincides with $\tilde{r}_0\doteq r_0$ . 
With the increasing local phase space density 
it slightly decreases. The maximal reduction factor of $1/\sqrt{2}$ 
is achieved for large ($r_0\Delta\gg 1$) and dense 
($\xi\rightarrow 1$, ${\bf p}\rightarrow 0$) systems. 
Considering the limit $q\rightarrow 0$ and  $r_0\Delta\gg 1$, 
we can neglect the $q$-dependence of the condensate term 
($r_{{\rm c}}\ll r_{{\rm g}}$) and of the product of 
the single-particle spectra 
in the denominator of the correlation function and, using 
Eqs. (\ref{cfinc}) and (\ref{Gapp}), write 
\begin{equation} 
R({\bf p}_1,{\bf p}_2)\doteq 1+\exp\left(-\frac{ 
N_{{\rm g}}({\bf p})} 
{N_{{\rm g}}({\bf p})+N_{{\rm c}}({\bf p})}r_{{\rm g}}{}^2{\bf q}^2\right). 
\label{cfincapp<} 
\end{equation} 
It follows from Eq. (\ref{cfincapp<}) that the condensate contribution 
leads to an additional reduction of the interferometry radius squared 
(defined as a low-${\bf q}^2$ slope of the correlation function) as 
compared with the case of a pure BE gas. In case of a dominant 
BE condensate the interferometry radius tends to zero whatever 
large is the geometric size of the system. 
 
Note however that, due to the non-Gaussian character of the correlation 
functions at large phase space densities, their real width is determined 
by the large-${\bf q}^2$ behavior which reads as: 
\begin{equation} 
R({\bf p}_1,{\bf p}_2)\doteq 1+\left(\frac{ 
N_{{\rm c}}({\bf p})} 
{N_{{\rm g}}({\bf p})}\right)^2\exp(-r_{{\rm c}}{}^2{\bf q}^2). 
\label{cfincapp>} 
\end{equation} 
Requiring $R({\bf q}_{{\rm eff}})=1+1/{\rm e}$, we get for the corresponding 
effective radius squared: 
\begin{equation} 
r_{{\rm eff}}{}^2\equiv {\bf q}_{{\rm eff}}{}^{-2}=r_{{\rm c}}{}^2\left/\right. 
\{1+2\ln[N_{{\rm c}}({\bf p})\left/\right.N_{{\rm g}}({\bf p})]\}. 
\label{reff} 
\end{equation} 
Thus, compared with the vanishing 
of the low-${\bf q}^2$ slope of the correlation function (as $(1-\xi)$), 
the effective radius squared vanishes at large phase space densities 
much slower (as $1/|\ln(1-\xi)|$) (see Fig. \ref{CFI}). 
 
It should be noted that Eqs. (\ref{n_inc}) and (\ref{cfinc}) assume that 
the initial Poissonian multiplicity distribution extends to any arbitrarily 
large number of bosons. In reality, however, this number is limited 
due to the finite available energy 
(for a study of the energy constraint effect on pion multiplicity 
see second paper in ref. \cite{cha95}). 
It is therefore interesting to see 
how fast the semi-inclusive spectra 
approach the inclusive limit with the increasing number 
$n_{\rm max}$ of the included pions. In Fig. 9 we demonstrate the 
$n_{\rm max}$-dependence of the semi-inclusive correlation functions 
for a fixed value of the density parameter $\xi=0.95$ and, 
in Fig. 10 - the $n_{\rm max}$-dependence of the 
correlation function intercepts 
for different $\xi$-values. 
We can see that the width of the semi-inclusive correlation function 
increases with the increasing $n_{\rm max}$, while its intercept decreases 
at small $n_{\rm max}$, reaching a minimum at $n_{\rm max}\approx \langle n\rangle$, 
and then approaches the limiting value of 2 roughly as $\log n_{\rm max}$. 
The inclusive behavior is practically saturated at 
a moderate number of the included pions 
$n_{\rm max}= k\langle n\rangle$, 
where $k$ increases with the density parameter $\xi$ 
from about 3 at $\xi=0.89$ to about 5 at $\xi=0.99$.\footnote 
{The increase of the saturation point $n_{\rm max}/\langle n\rangle$ 
with the density is related with the increasing condensate 
contribution which, for the ideal BE gas, 
is characterized by very large multiplicity fluctuations. 
} 
Thus the neglect of the 
energy-momentum constraints 
in Eqs. (\ref{n_inc}) and (\ref{cfinc}) can be justified 
provided $\langle n\rangle \ll \sqrt{s}/m$, {\it e.g.}, in the usual case of a 
logarithmic increase of the mean multiplicity with the c.m.s. energy $\sqrt{s}$. 
The situation can change in the case of very large and dense systems dominated 
by a soft condensate. Then the regime $\langle n\rangle \sim \sqrt{s}$ can 
settle, the energy-momentum constraints leading to the reduction of the 
maximal effective number of produced pions to 
$n_{\rm max} \sim\langle n\rangle $ and, as a result, to the suppression of the 
measured inclusive correlation function. Clearly, such an eventual suppression 
has nothing in common with the coherence effect.

\section{Discussion and conclusion}\label{sec6} 
 
We have illustrated an approximate scaling of 
multiboson effects with the density parameters $\xi, \xi_n, \dots$ 
(see, {\it e.g.}, Figs. \ref{NPM1} and \ref{ratsp}). It means that 
though our numerical results were obtained for typical AGS or SPS 
multiplicities 
of the order of tens or hundreds of pions, they are 
approximately valid also for 
higher multiplicities expected at RHIC or LHC 
energies. 
 
The value of the density parameter $\xi$ can be estimated with the help of 
Eq. (\ref{8.51aa}) relating the phase space density 
in the rare gas limit with the 
integrated correlation function. Thus 
using the usual Gaussian parameterization for the correlation function in the 
longitudinally co-moving system (LCMS): 
\begin{equation} 
R(p_1,p_2)=1+\lambda \exp(-r_x^2q_x^2-r_y^2q_y^2-r_z^2q_z^2), 
\label{8.2} 
\end{equation} 
where $x$, $y$ (${\bf y} \parallel {\bf z}\times {\bf p}$) 
and $z$ denote the outward, sideward and longitudinal 
directions respectively and parameterizing 
the single-particle spectra as 
\begin{equation} 
{N}({\bf p})=\frac{dn}{dy} 
\frac{\exp(-(m_t-m)/T)}{2\pi T(T+m)m_t\cosh y}, 
\label{8.3} 
\end{equation} 
we arrive at the mean pion phase-space density 
\begin{equation} 
\langle f\rangle _{{\bf p}}= 
\frac{\lambda\pi^{3/2}}{V}N({\bf p})\cosh y 
=\lambda\frac{\sqrt\pi}{2} 
\frac{\exp(-(m_t-m)/T)}{VT(T+m)m_t}\frac{dn}{dy}, 
\label{8.4} 
\end{equation} 
where $V=r_xr_yr_z$ is the LCMS interference volume.\footnote 
{A better estimate may require the substitution  \cite{bar97} 
$\lambda \rightarrow \lambda^{1/2}$ in Eq. (\ref{8.4}). 
} 
For soft pions ($p_t\approx 0$ and $y\approx 0$) 
at SPS energies this quantity is typically $\sim 0.2$. 
Since this value is sufficiently small, we can compare it with 
the model phase space density in the rare gas limit 
$\langle \tilde{f} \rangle_{{\bf p}=0} \approx \eta/(\sqrt{2}r_0\Delta)^3$ 
(see Eq. (\ref{8.51a})) and get 
$\xi\approx 0.4-0.5$. 
For such values of the density parameter our calculations 
point to rather small multiboson effects in the ordinary events. 
These effects can show up however 
in the events containing sufficiently high density fluctuations. 
Particularly, the condensate effects could be seen 
in certain high multiplicity events (see, {\it e.g.}, Fig. \ref{intercept}) 
in which the phase space volume or subvolume 
$\sim (r_0\Delta)^3$ does not follow 
the increasing multiplicity (as it presumably does in the ordinary events) and 
remains 
sufficiently small to guarantee a nonvanishing factor $\xi^{r_0\Delta}$ 
determining the condensate size (see Eqs. (\ref{n-appr}) and 
(\ref{Pasyincpar})). 
 
Since 
at present energies the LCMS interference volume $V$ seems to 
scale with $dn/dy$, the freeze-out of the pions occurs on average 
at approximately constant phase space density (see Eq. (\ref{8.4})). 
In the rare gas regime, 
based on the density scaling one can then expect 
about the same relative size of the multiboson effects also at RHIC and 
LHC energies, up to 
a slight increase in $\xi$ due to the vanishing of the finite-size 
corrections with the increasing phase space volume of the 
emitting system. 
At the same time, 
the growing phase space volume will lead to suppression of 
the average condensate contribution, 
determined by the factor $\xi^{r_0\Delta}$. 
 
Considering the multiboson effects in the low (BE gas) and the large 
(BE condensate) density limits, we have obtained simple 
analytical formulae 
accounting for the finite size of the phase space volume and 
allowing to follow the dependence of the mean 
multiplicities, single-boson spectra and two-boson correlation 
functions on the phase space density parameters. 
In principle, these formulae provide a possibility to 
identify multiboson effects among others. Particularly, 
the width of the low-$p_t$ enhancement 
due to the BE condensation decreases with the size of the 
system as $r_0^{-1/2}$ and this narrowing makes the observation of the 
effect easier. 
 
The results of the considered simple model should not be taken, however, 
too literally since: 
 
a) due to its static character, the model does not explicitly account for 
the experimental indications on 
a constant freeze-out phase space density and the related expansion 
of the emission volume. 
The qualitative application of our model to heavy ion collisions 
is however possible in the limited phase space regions. For example, 
the pions with a rapidity 
difference greater than about unity have to be considered 
as originating from different static sources; 
As discussed in Section \ref{sec2b}, the residual slow relative motion, 
if decoupled from other source characteristics, 
merely leads to a wider momentum dependence of the emission 
function in Eq. (\ref{8.71}): 
$\Delta^2 \rightarrow \Delta^2+\Delta_0^2$; 
 
b) due to a mixture of different production processes 
({\it e.g.}, due to contribution of different impact parameters), 
the real multiplicity distribution and particle spectra will be 
rather weighted sums of those in Eqs. (\ref{mult1})-(\ref{8.13}) 
calculated with different sets of the parameters 
$\eta^{i}, r_0^{i}, \Delta^{i}$. As a result, near the condensate 
limit, the multiplicity distribution can be wider than the BE one 
and the intercepts of the inclusive and fixed-$n$ correlation 
functions can differ from the respective single--process values 
of 2 and 1; 
 
c) when estimating the freeze-out phase space density 
from the experimental data, the 
multiboson system is considered as a homogeneous medium. 
However, there can be large local density fluctuations - speckles 
which can give rise to 
noticeable multiboson effects even at a moderate value 
of the mean phase space density; 
 
d) on the other hand, the multiboson effects can be somewhat 
suppressed due to a possible violation of the factorization 
assumption in Eq. (\ref{8.74a}) or due to the lack of the reflection 
symmetry of the emission volume. In latter case the functions 
$G_n({\bf p}_1,{\bf p}_2)$ are no more real; 
 
e) for identical charged pions, the BE effects are also suppressed 
due to the Coulomb repulsion. Since this repulsion is important only 
in a weakly populated region of very small relative momenta 
determined by the pair Bohr radius $a=387$ fm, 
the suppression of the global BE weights $\omega_n$ is rather small. 
For example, for $\omega_2$ this suppression, being about 
$(a r_0\Delta^2)^{-1}$, is usually less than one per mill. 
The Coulomb distortion of the global multiboson effects is 
therefore negligible in the rare gas limit. 
Nevertheless, since 
the Coulomb repulsion destroys the formation of the condensates 
made up from positive and negative pions in the disjoint 
phase space regions, it can lead to noticeable differences 
between charged and neutral pions 
in dense systems. 
Particularly, we can expect a decrease of the 
charge-to-neutral multiplicity ratio with the increasing 
phase space density. 
 
Because of large numbers of positive and negative pions produced 
in heavy ion collisions, 
one could also raise a question about importance of the 
Coulomb screening effects violating standard two-body 
treatment of the correlations in the low density limit. 
There are however arguments 
showing that the screening will be of minor importance even at LHC 
\cite{alinote95} (see also \cite{anch96}). 
Note that in the 
scenario with a constant phase-space density the corresponding Debye radius 
\begin{equation} 
r_D=[4\pi (\rho_++\rho_-)e^2/T]^{-1/2}, 
\label{8.52} 
\end{equation} 
where $e^2=1/137$ and $\rho_++\rho_-$ 
is the total density of charged pions in the 
configuration space, will be also constant, up to a weak energy 
dependence due to the temperature T. Assuming that pions with a 
rapidity difference greater than unity come from spatially disjoint regions 
of phase-space, we can put \cite{alinote95} 
\begin{equation} 
\rho_+=\left(\frac{\lambda}{(2\pi )^3}\right)^{1/2} 
\frac{dn_+/dy}{V}= 
\frac{\sqrt 2}{\sqrt \lambda \pi^2}T^3\langle f_+\rangle _{y} 
\label{8.62} 
\end{equation} 
and obtain $r_D\approx 15$ fm at $\langle f_+\rangle _{y}\approx 0.1$ and 
$T\approx 200$ MeV/c ($r_D\sim 1/T$). Thus at LHC energies we can expect the 
characteristic distances between the pion production points 
comparable or larger than the screening radius $r_D$ leading to a 
suppression of the usual two-particle Coulomb effects. 
In fact, two charged pions produced at a distance $r^* > r_D$ 
start to feel their Coulomb field only after some time when the density 
decreases to a value corresponding to Debye radius larger than $r^*$. 
During this time the vector of the relative distance between the pion emission 
points increases by \cite{alinote95} 
\begin{equation} 
\Delta {\bf r}^* \sim \frac{{\bf k}^*}{(mT)^{1/2}}V^{1/3} 
[\left(\frac{r^*}{r_D}\right)^{2/3}-1]. 
\label{8.74} 
\end{equation} 
Substituting ${\bf r}^*$ by ${\bf r}^*+\Delta {\bf r}^*$ 
in the argument of the Coulomb wave function, we can see however that the 
suppression of the Coulomb effect can be substantial only in the 
region of large relative momenta $k^* > (mT)^{1/2}$ 
where the correlations due to QS 
and FSI are already negligible (see, {\it e.g.}, \cite{ll90,ll82}). 
 
Finally, to make easier the understanding of our results 
in context of other papers and, 
for the reader's convenience, we compare our results 
(including those in \cite{alinote95}) with the results of refs. 
\cite{pra93,pra94,cha95,zim97} and the recent papers 
\cite{zha97,wie98,bia98,ray98,mis97,zsh98,bz99b,sal98,mek98,bia99} 
which appeared either after the present 
work was basically completed or during the process of its evaluation. 
 
In the pioneering papers \cite{pra93,pra94}, the analytically solvable 
model discussed in Section \ref{sec4} was introduced. A simple algorithm was 
given allowing to calculate BE modified multiplicity distribution, 
single- and two-boson spectra in terms of the quantities 
$C_n^{Pratt}=\widetilde{g}_n/n$, 
$G_n^{Pratt}({\bf p}_1,{\bf p}_2)  = 
\widetilde{G}_n({\bf p}_1,{\bf p}_2)$. 
The details of the calculation of the $G_n$-functions and of the 
two-boson spectra were not given in \cite{pra93,pra94}. They can be 
found in \cite{alinote95,cha95,zim97,ame-led} (see also \cite{zha97,wie98}). 
In refs. \cite{zha97} this technique was extended to the spectra 
and correlations of three or more pions. It was shown that, in the 
inclusive case, the old formalism, not accounting for the multiboson 
effects, can be recovered by a redefinition of the Wigner-like density 
which then becomes more narrow both in momentum and configuration space. 
 
Regarding the choice of the Wigner-like density $D(p,x)$, 
in the original papers \cite{pra93,pra94}   and in 
\cite{cha95}, it was slightly different from that 
in Eq. (\ref{8.71}), corresponding to the substitution 
$\exp(-{\bf q}^2/2\Delta^2) \rightarrow \exp(-p_0/T)$, where 
$T=\Delta^2/m$. This choice is, in fact, in contradiction with 
the uncertainty principle, the latter requiring an energy independent 
density in the considered case of a fixed emission time $t=0$ 
(see Eq. (\ref{21A})). (For the same reason, the freedom in the choice 
of the parameters $r_0$, $\Delta$ is limited by the inequality 
$r_0\Delta \ge 1/2$, the equality corresponding to the zero distance 
between the emitter centers.) 
The incorrect choice of the Wigner-like density in 
\cite{pra93,pra94,cha95} can be cured by the 
substitution $r_0^2\rightarrow r_0^2-(2\Delta)^{-2}$. 
To recover the recurrence relations in Eqs. (\ref{8.73}), 
besides this substitution, one has to take into account 
that the width and the slope parameters used {\it e.g.} in 
\cite{cha95} are related to ours as: 
$R_0^2=2r_0^2, p_0^2=\Delta^2, a_n=b_n^++b_n^-, g_n=2(b_n^+-b_n^-)$ 
and that a factor of $R_0^2$ is missing in Eq. (66) for $g_{n+1}$ 
in \cite{cha95}. 
 
The analytical solution for 
the $G_n$-functions has been found in \cite{zim97,wie98,bia98}. 
In \cite{zim97} the same form of the Wigner-like density as in 
Eq. (\ref{8.71}) was used with the parameters $R_{\rm eff}^2=r_0^2$ and 
$\sigma_T^2=2\Delta^2$. However, since the analytical solution 
was derived based on the recurrence relations of ref. \cite{cha95}, 
it has also to be cured by the substitution 
$r_0^2\rightarrow r_0^2-(2\Delta)^{-2}$. 
Particularly, after this substitution, Eq. (200) in the preprint 
version of \cite{zim97} for the critical multiplicity $\eta_c$ 
(coinciding with Eq. (9) in \cite{pra93} corrected for the 
misprints \cite{zim97}) then reduces to the simple result 
\cite{alinote95}: $\eta_c=\beta$. 
In \cite{wie98}, the analytical solution given in Eqs. (2.14) 
corresponds to the emission function in Eq. (\ref{8.71}) with the 
parameters $r_0^2=R^2/2+(2p_0)^{-2}$, $\Delta^2=p_0^2$.\footnote{ 
Eqs. (2.14b) and (2.14d) have to be corrected for the misprints 
by the substitutions: 
$1/4\rightarrow (1+c/2)/4$ and $(1+c)\rightarrow (1+c/2)$ respectively. 
Note that the other solution given in Eqs. (4.3), (4.4) of ref. 
\cite{wie98} presumably contains an error since, corresponding 
to a Gaussian factorizable model, it does not reduce to the general 
solution in Eqs. (\ref{8.72}-\ref{exact}) (particularly, it shows no 
condensate behavior at high densities). 
} 
Requiring a matrix algebra, this solution is however less transparent 
compared with that in \cite{zim97}. 
In \cite{bia98}, the analytical solution was obtained 
using the technique of the density matrix in the 
1-dimensional momentum space. Generalizing this solution to three 
dimensions (by substituting the normalization factors 
$\lambda_0^n/[1-(1-\lambda_0)^n]$ and $(2\pi\hat{\Delta}_n^2)^{-1/2}$ 
by their cubes) and relating the notation of ref. \cite{bia98} 
to ours: $\hat{\Delta}_n^{-2}=8b_n^+$, 
$\hat{R}_n^2=2b_n^-$, $\omega_n^{BZ} =(1-\lambda_0)^n=\rho^n$, 
$L(p_1,p_2)=\widetilde{G}_n(p_1,p_2)$, 
one can see that it then coincides with the solution in 
Eqs. (\ref{exact}). 
 
As for the analytical approximations of the mean multiplicity, 
single- and two-boson inclusive spectra, similar results as in 
ref. \cite{alinote95} have been obtained in \cite{zim97} in 
both the low and high density regimes. The behavior of the 
single-boson spectrum in the low density limit 
was also studied numerically 
in \cite{ray98} - a linear increase of the relative correction 
with the phase space density was found (in agreement with 
Eq. (\ref{Pasyincpar}) at small $\widetilde{\xi}_{\bf p}$). 
The condensate behavior at very high densities was also 
obtained in \cite{bia98}. 
A comment requires 
the effective radius squared, estimated as \cite{bia98} 
$r_{\rm eff}^2=(2\langle {\bf q}^2\rangle)^{-1}$, where the averaging 
over the correlation term is assumed. With the increasing mean 
multiplicity, $r_{\rm eff}^2$ decreases from 
$r_0^2$ to $r_0^2/(2r_0\Delta)$. Note, however, that at high 
densities, $r_{\rm eff}^2$ has little to do with the interferometry 
radius squared. The latter accounts also for the change of the 
single-boson spectrum and, with the increasing density, 
decreases from $r_0^2-(2\Delta)^{-2}$ to 0. 
 
A special comment requires the normalization of the correlation 
functions. There are two popular definitions of the two-particle 
correlation function, both representing a ratio of the two-particle 
spectrum to the product of the single-particle ones, 
but differing in the normalization. In case (I) the spectra are 
normalized to 1 while in case (II) - to the numbers of single- 
and two-particle counts. This definitions correspond to 
Eq. (\ref{cf2}) or (\ref{cf2i}) with (I) $c_n=n/(n-1)$ or 
$c=\langle n\rangle^2/\langle n(n-1)\rangle$ and (II) $c_n=c=1$. 
In the inclusive case, based on the thermal-type models 
(see the comment after Eq. (\ref{cfinc})), the second choice was 
advocated \cite{mis97,zsh98}. 
Recall that the first choice (I) would be preferable in case 
of a fixed pion multiplicity provided a large phase space volume 
(see discussion after Eq. (\ref{cn})). 
Generally, however, there is no well-defined normalization and, 
for a reliable comparison with the experimental correlation functions, 
a free normalization parameter, depending on the production 
mechanism (not necessarily a single thermal one) and experimental 
conditions, has to be introduced  \cite{alinote95}. 
 
The role of the normalization was misunderstood in \cite{cha95,zim97}, 
where the decrease of the intercept of the inclusive correlation 
function (I) with the increasing phase space density, 
obtained in the factorizable Gaussian model, 
was incorrectly interpreted as a coherent laser behavior. 
Recall that pions in a coherent state would have a narrow Poisson 
multiplicity distribution, while the BE condensate is characterized by 
very wide multiplicity fluctuations. Clearly, the coherent pion 
production requires a special mechanism (not present in the 
model), like that leading to possible formation of the disoriented 
chiral condensate - DCC (for a review, see \cite{bjo97}). 
Possibilities of experimental investigations of BE condensate and 
DCC phenomena have been recently discussed in \cite{bz99b}. 
A discussion of statistical physics aspects of the multiboson effects 
can be found in \cite{pra93,zim97,wie98,zsh98,sal98,mek98,bia99}. 
 
In conclusion we summarize the results. 
 
%\begin{itemize} 
%\item 
- {The influence of the multiboson effects on boson multiplicities, 
single-boson spectra and two-boson correlations, 
including an approximate scaling behavior 
of some of their characteristics with the 
phase space density ({\it e.g.}, $\langle n\rangle/\eta$ {\it vs} $\xi$ or
$N_n(0)/\beta_n$ {\it vs} $\xi_n$), 
has been demonstrated 
using the analytically solvable Gaussian model.} 
 
%\item 
- {The approximate analytical formulae are given 
allowing to follow the dependence of these 
quantities on the phase space density parameters 
thus providing a possibility for the 
identification of the multiboson effects among others.} 
 
%\item 
- {The meaning and the applicability conditions of the model 
factorization assumption are clarified using the physically transparent 
Kopylov--Podgoretsky {\it ansatz} of classical one-particle sources. 
For heavy ion collisions, the factorizatiom 
assumption is expected to be valid 
in case of an impact parameter selection.} 
 
%\item 
- {The lowest order cumulant approximation, suggested for a practical 
account of multiboson effects in realistic transport code 
simulations \cite{ame-led}, has been shown to be reasonable 
at moderate densities indicated by the experimental data.} 
 
%\item 
- {At high densities, the spectra are mainly determined by the 
universal condensate term $P_c({\bf p})$ ({\it e.g.}, 
$N_n^{(1)}({\bf p})\rightarrow n P_c({\bf p})$ and 
$N_n^{(2)}({\bf p}_1,{\bf p}_2)\rightarrow n(n-1) 
P_c({\bf p}_1)P_c({\bf p}_2)$) and the initially narrow Poisson 
multiplicity distribution approaches a wide BE one: 
$\langle n(n-1)\rangle\rightarrow 2\langle n\rangle^2$. 
As a result, the intercepts of the inclusive and fixed-$n$ 
correlation functions (properly normalized to 1 at large $|q|$) 
approach 2 and 1, respectively and their low-${\bf q}^2$ slopes rapidly 
vanish with increasing density; the corresponding increase of the 
apparent correlation function width is however rather slow - 
logarithmic in the density.} 
 
%\item 
- {It is found that, even near the condensate regime, the inclusive 
characteristics saturate at rather moderate multiplicities of some 
multiples of $\langle n\rangle$ thus justifying the neglect 
of energy--momentum constraints in the considered analytical model. 
The latter are however important near a multipion threshold, 
particularly making impossible the production of a very 
cold BE condensate.} 
 
%\item 
- {Though spectacular multiboson effects are hardly to be 
expected in typical events 
of heavy ion collisions in present and perhaps also in future 
heavy ion experiments, they can clearly show up in certain classes 
of events containing sufficiently high density fluctuations.} 
%\end{itemize} 
 
\section*{Acknowledgments} 
This work was supported in part by GA Czech Republic, 
Grant No. 202/98/1283,
by Russian Foundation of Fundamental Investigations, 
Grant No. 97-02-16699,
by Ukrainian State Fund for the Fundemental Research,
Contract No. 2.5.1/057 and
by Ukrainian--Hungarian Grant No. 2M/125-199. 
 
%\begin{thebibliography}{99} 
 
%\newpage 
 
\begin{figure} 
\epsfig{file=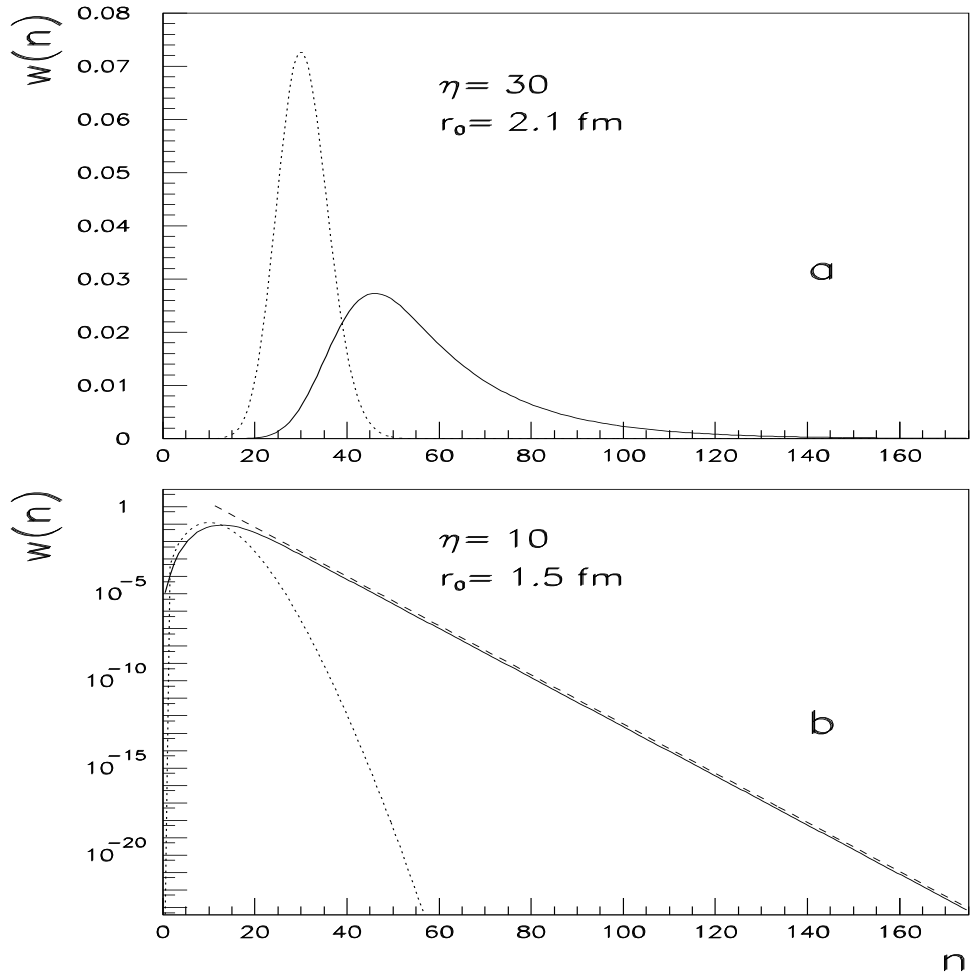, width=15cm, height=15cm} 
\caption{ The multiplicity distribution of neutral pions 
for a) $\Delta=0.25$ GeV/c, $r_0=2.1$ fm, $\eta=30$ and 
b) $\Delta=0.25$ GeV/c, $r_0=1.5$ fm, $\eta=10$, 
where $\eta$ is the mean multiplicity of the initial 
Poissonian distributions (dotted curves).
The dashed line corresponds to Eq. (\ref{mult}).} 
\label{NPM} 
\end{figure} 
 
\begin{figure} 
\epsfig{file=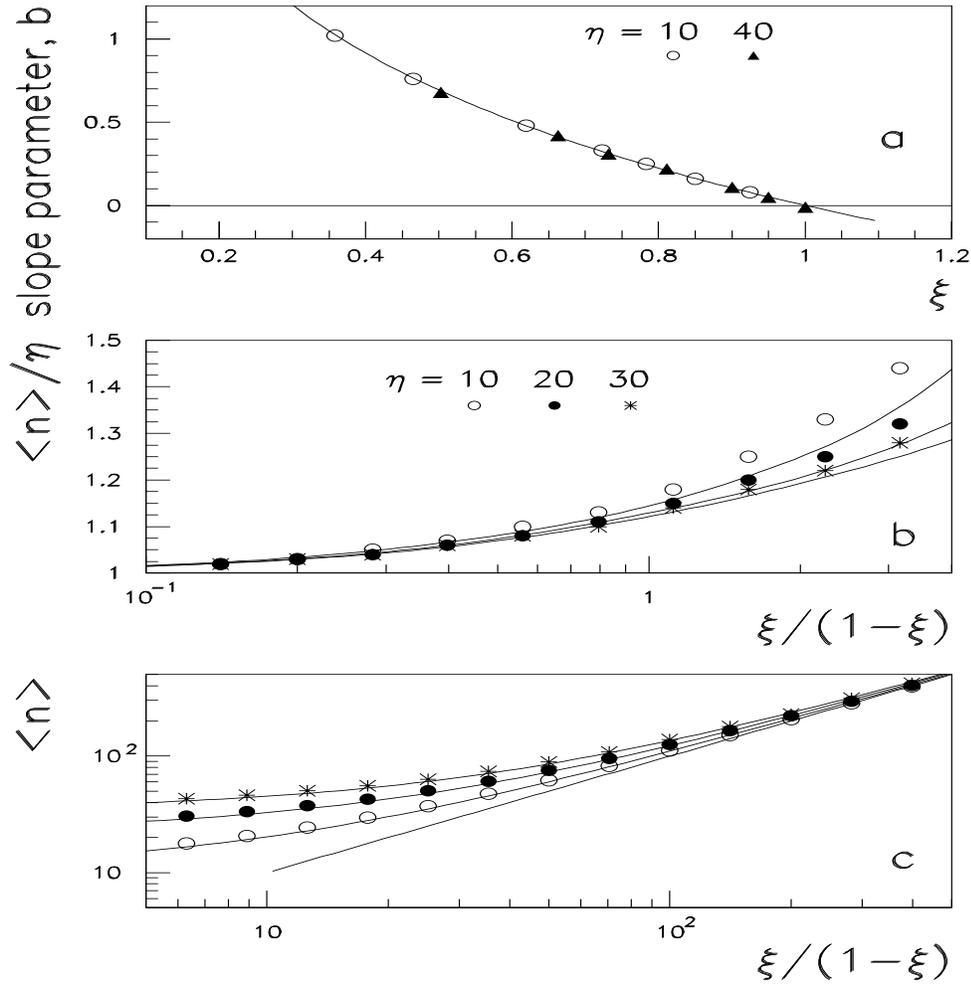, width=15cm, height=15cm} 
\caption{ The slope parameter $b$ of the exponential tail 
$c\cdot\exp(-b\cdot n)$ of the multiplicity distribution (a), 
the ratio of the mean multiplicity to the initial Poissonian one (b) 
and the mean multiplicity (c) 
as functions of the density parameters $\xi=\eta/\beta$ and $\xi/(1-\xi)$; 
$\Delta=$ 0.25 GeV/c. 
The curve in (a): $b=-\ln\xi$, the curves in (b), (c) are calculated 
according to the tailing approximation in Eq. (\ref{n-appr}), 
the line in (c): $\langle n\rangle=\xi/(1-\xi)\equiv 
\nu$.} 
\label{NPM1} 
\end{figure} 
 
\begin{figure} 
\epsfig{file=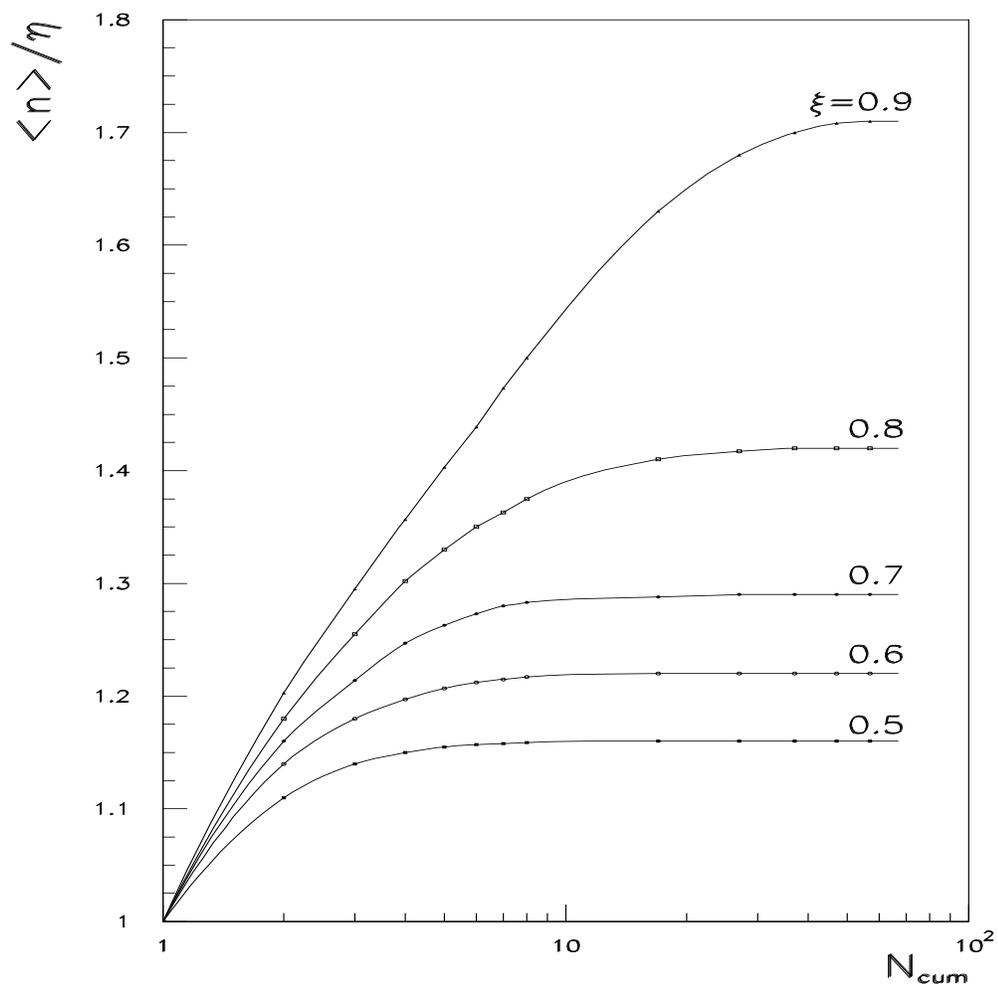, width=15cm, height=15cm} 
\caption{The ratio of the mean multiplicity to the initial Poissonian 
one as a function of the number of the contributing cumulants 
for different values of the density parameter $\xi$. 
} 
\label{Ncum} 
\end{figure} 
 
\begin{figure} 
\epsfig{file=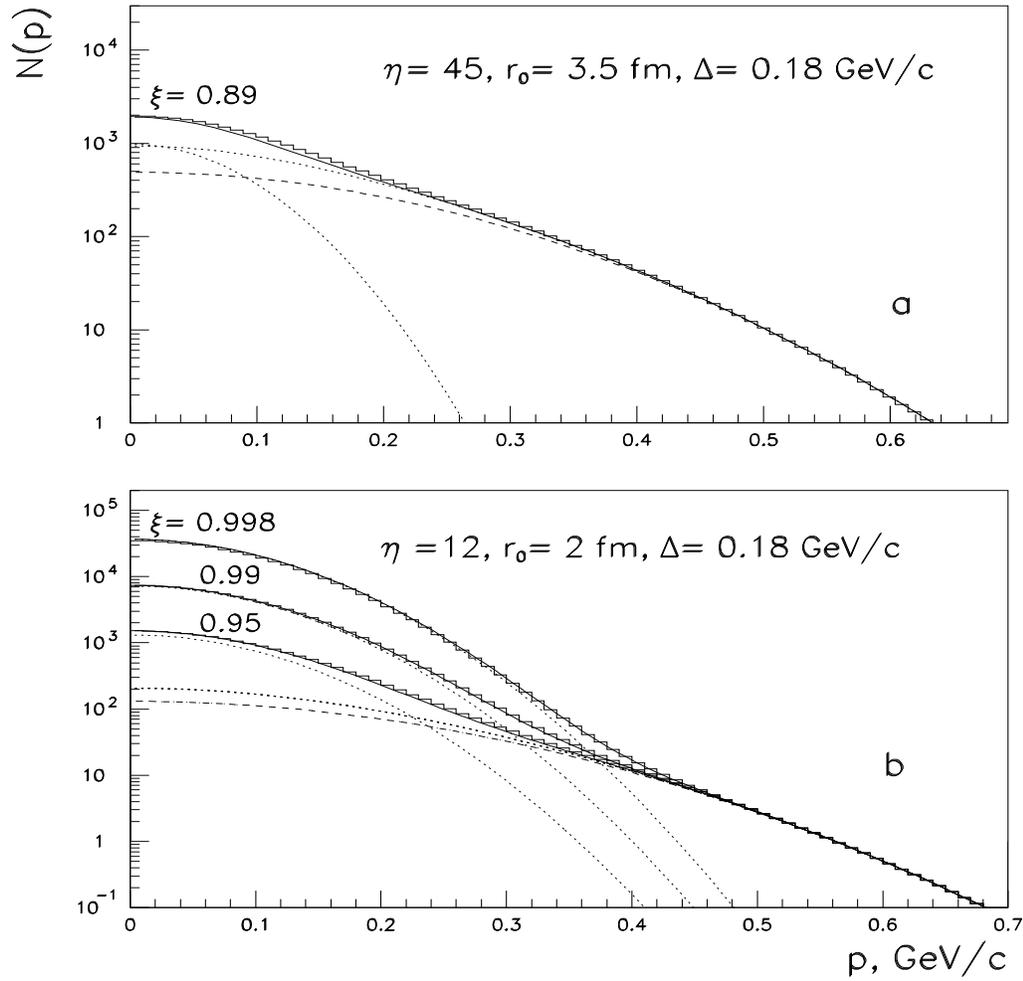, width=15cm, height=15cm} 
\caption{ The inclusive single-particle spectra corresponding to 
the density parameters 
a) $\xi=0.89$ and 
b) $\xi=0.95$, 0.99 and 0.998 
(the radius $r_0$ slightly varies near 2 fm);
the corresponding mean multiplicities are a) 64.3 and 
b) 33.5, 113.7 and 433.8. 
The histograms represent the exact result, the full 
curves are calculated according to the tailing 
approximation (\ref{Pasyincpar}), 
the dotted ones represent the contributions of the two 
(BE gas and BE condensate) terms in Eq. (\ref{Pasyincpar}) and 
the dashed curves correspond to the rare gas limit 
$\eta P({\bf p})$. 
} 
\label{SPS} 
\end{figure} 
 
\begin{figure} 
\epsfig{file=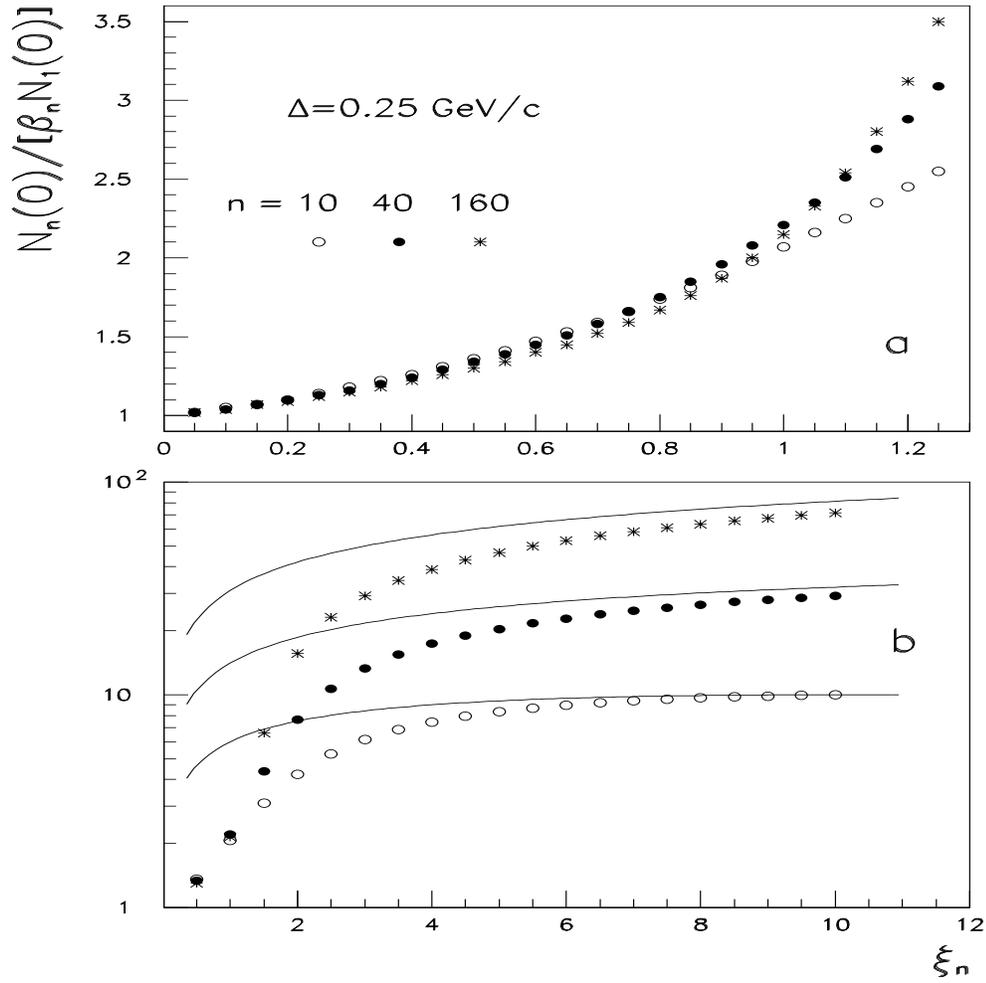, width=15cm, height=15cm} 
\caption{ The ratios of the BE 
affected single-particle spectrum to the dominant large-${\bf p}$ 
contribution 
$\beta_n {N}_1({\bf p})$, $\beta_n =n\omega_{n-1}/\omega_n$, 
calculated at ${\bf p}=0$ 
as functions of the density parameter $\xi_n =n/\beta$. 
The curves represent the large-$\xi_n$ limit: 
$(2r_0\Delta)^{3/2}\xi_n\equiv [2(n/\xi_n)^{1/3}-1]^{3/2}\xi_n$.} 
\label{ratsp} 
\end{figure} 
 
\begin{figure} 
\epsfig{file=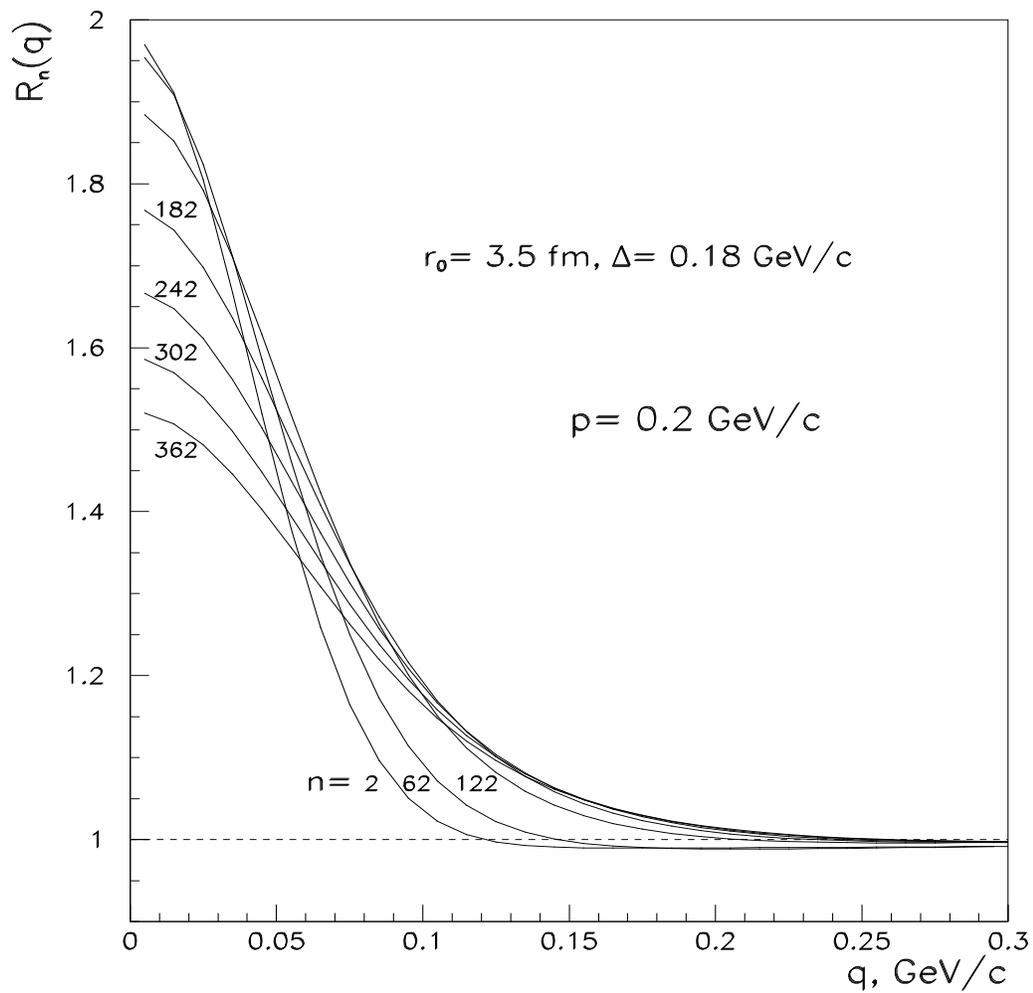, width=15cm, height=15cm} 
\caption{ The two-pion correlation functions for the 
multiplicities increasing from $n=$ 2 to 362 with a step of 60 
(the corresponding density parameter $\xi_n$ ranging from 0.04 
to 7.2 with a step of 1.2). 
The higher is the multiplicity the lower is the 
intercept of the correlation function and the larger is its width.} 
\label{CFN} 
\end{figure} 
 
\begin{figure} 
\epsfig{file=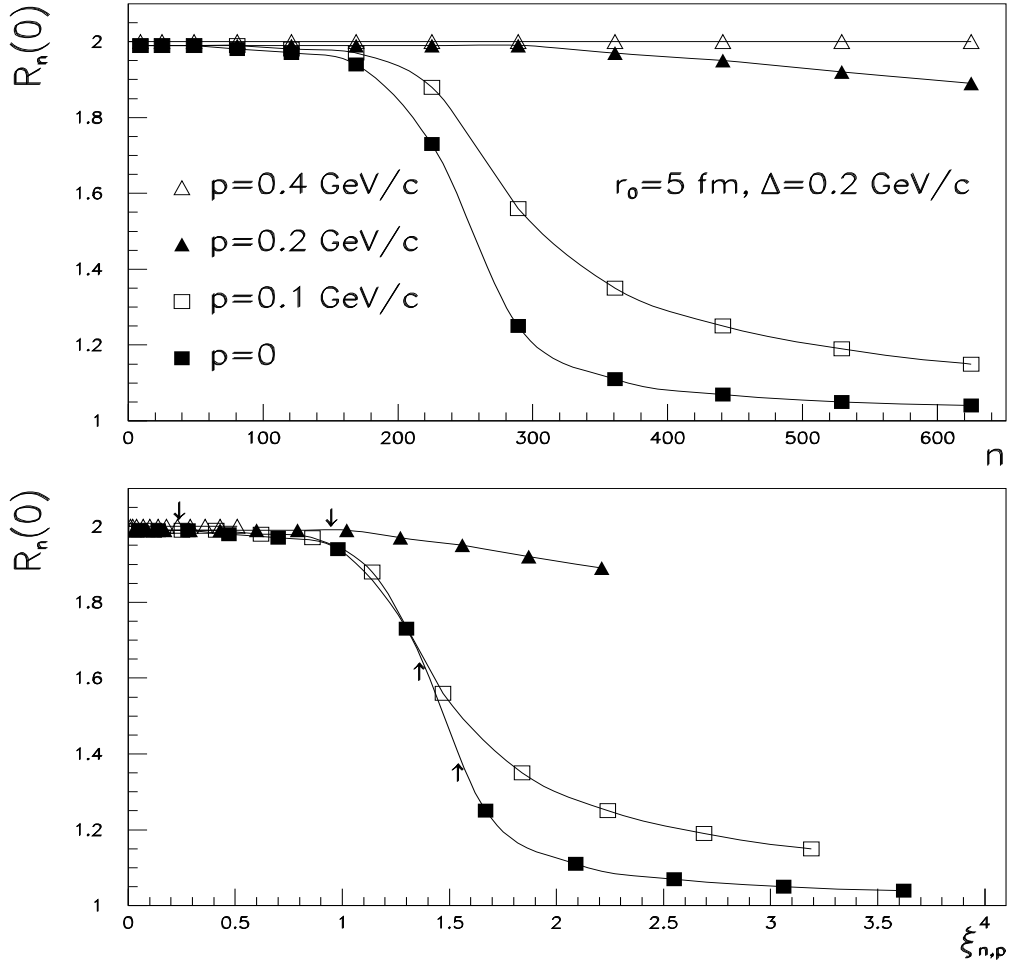, width=15cm, height=15cm} 
\caption{ The intercept of the two-pion correlation functions 
as a function of the multiplicity $n$ 
and the density 
parameter $\xi_{n,{\bf p}}$ for several values 
$p=0$, 0.1, 0.2 and 0.4 GeV/c 
of the mean momentum of the two pions. 
The arrows on the interpolating curves 
indicate the intercept values corresponding to 
$\xi_n=3\xi=1.5$ ($n\approx 3\langle n\rangle$).} 
\label{intercept} 
\end{figure} 
 
\begin{figure} 
\epsfig{file=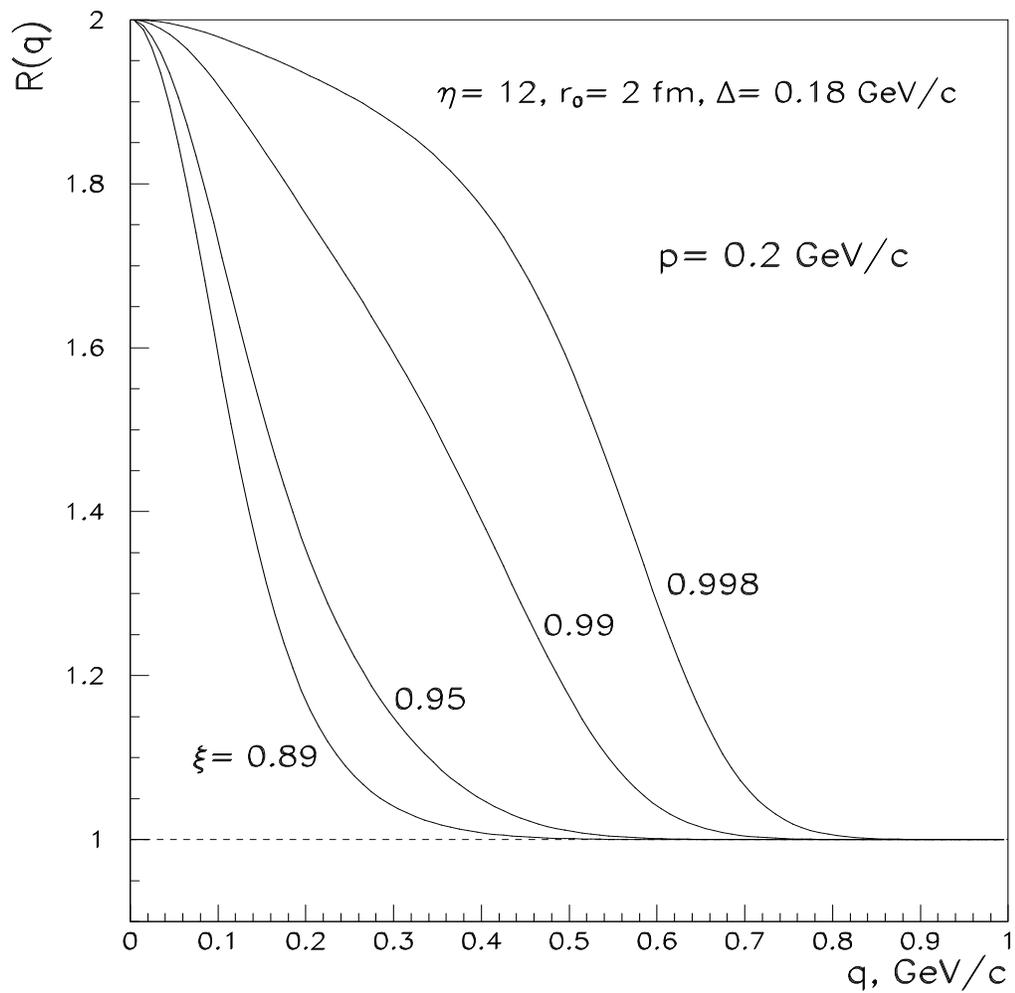, width=15cm, height=15cm} 
\caption{ The inclusive two-pion correlation functions 
demonstrating the increase of the correlation width with the 
increasing density parameter $\xi$. 
The different $\xi$-values are achieved by slight variations 
of the radius $r_0$ around 2 fm. 
} 
\label{CFI} 
\end{figure} 
 
\begin{figure} 
\epsfig{file=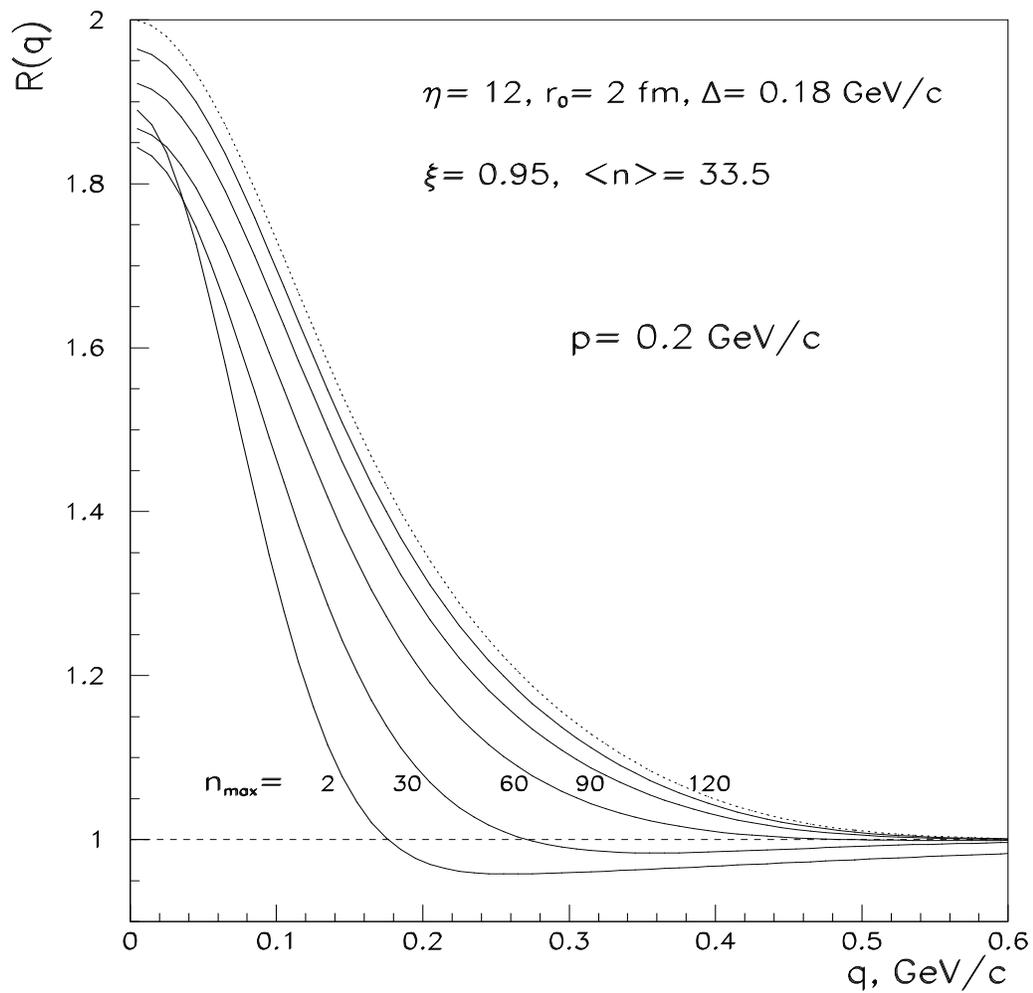, width=15cm, height=15cm} 
\caption{ The semi-inclusive 
correlation functions including the pion multiplicities from 0 to 
$n_{\rm max}$ for different values of $n_{\rm max}$. 
The dotted curve is the inclusive ($n_{\rm max}\rightarrow \infty$) 
correlation function. 
The conditions are the same as in Fig. \ref{CFI} 
for the density parameter $\xi=$ 0.95. } 
\label{CFSI} 
\end{figure} 
 
\begin{figure} 
\epsfig{file=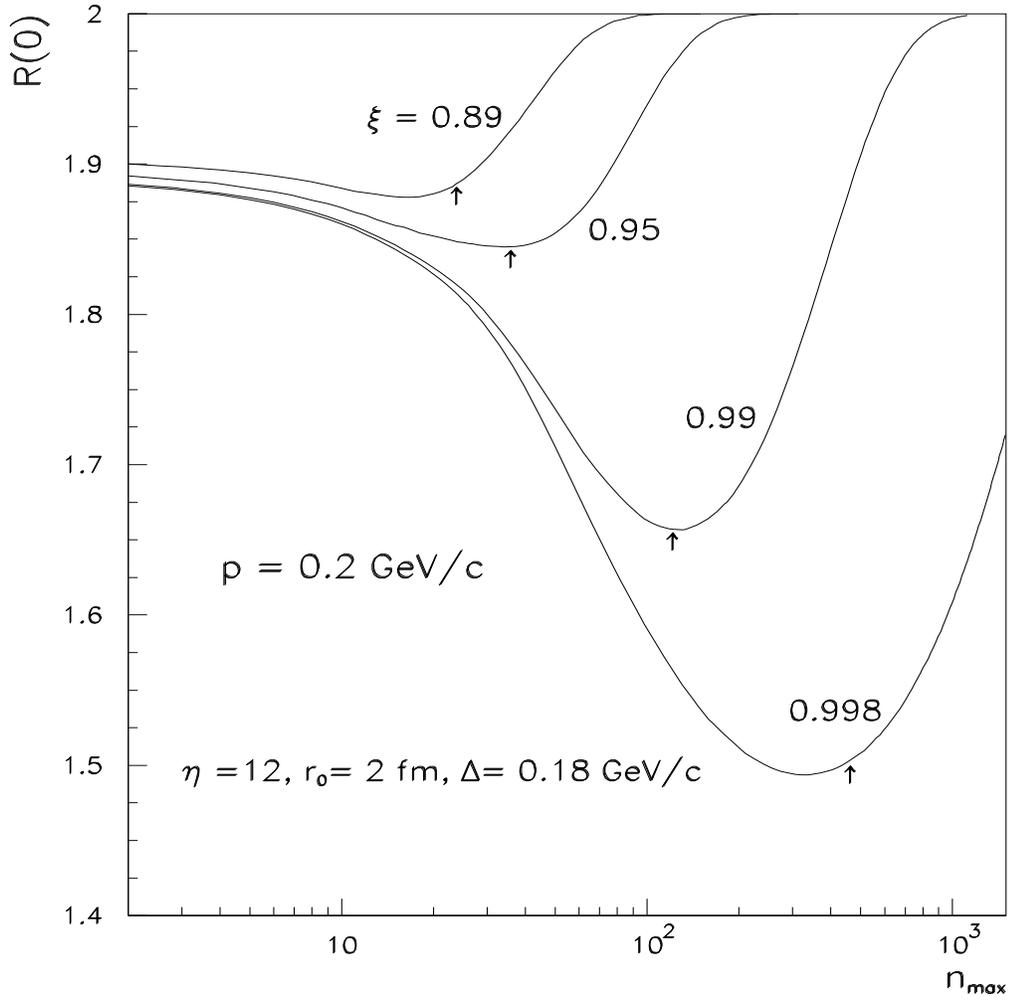, width=15cm, height=15cm} 
\caption{ The intercepts of the semi-inclusive 
correlation function including the pion multiplicities from 0 to 
$n_{\rm max}$ as functions of $n_{\rm max}$ for different values 
of the density parameter $\xi=$ 0.89, 0.95, 0.99 and 0.998; 
the arrows indicate the corresponding mean multiplicities 
$\langle n\rangle=$ 22.3, 33.5, 113.7 and 433.8. 
The conditions are the same as in Fig. \ref{CFI}. } 
\label{CFSI0} 
\end{figure} 
\end{document}